\documentclass[twocolumn,aps,pre,showpacs,floats,floatfix]{revtex4}

\usepackage{amssymb,amsmath}
\usepackage{graphics}
\usepackage{color}
\usepackage{epstopdf}
\usepackage{epsfig}
\usepackage{rotating}
\usepackage{float}
\usepackage{bm}
\usepackage{soul}
\usepackage{hyperref}
\usepackage{breakurl}
\usepackage[stable,perpage]{footmisc}
\usepackage{xr}
\newcommand{\rem}[1]{}
\DeclareMathAlphabet{\mathbi}{OML}{cmm}{b}{it} 
\newcommand{\non}{\nonumber}
\newtheorem{theorem}{Theorem}

\newcommand{\bx}{\mathbi{x}}

\newcommand{\bel}{\begin{equation}\label}
\newcommand{\ee}{\end{equation}}
\newcommand{\beq}{\begin{eqnarray}\label} 
\newcommand{\eeq}{\end{eqnarray}} 
\newcommand{\bc}{\begin{center}} 
\newcommand{\ec}{\end{center}} 
\newcommand{\ben}{\begin{enumerate}}
\newcommand{\een}{\end{enumerate}}
\newcommand{\bit}{\begin{itemize}}
\newcommand{\eit}{\end{itemize}}
\newcommand{\I}{\int_{\mathcal{V}}}

\newcommand{\bdf}{\mathbi{f}}

\newcommand{\bu}{\mbox{\boldmath$u$}}

\newcommand{\bom}{\mbox{\boldmath$\omega$}}

\newcommand{\bg}{\mbox{\boldmath$g$}}

\newcommand\shalf{\ensuremath{{\scriptstyle\frac{1}{2}}}}

\newcommand\quart{\ensuremath{{\scriptstyle\frac{1}{4}}}}

\newcommand\threehalves{\ensuremath{{\scriptstyle\frac{3}{2}}}}
\newcommand\fivehalves{\ensuremath{{\scriptstyle\frac{5}{2}}}}

\newcommand\onefifth{\ensuremath{{\scriptstyle\frac{1}{5}}}}

\begin{document}

\title{\color{red}A regularity criterion for solutions of the three-dimensional Cahn-Hilliard-Navier-Stokes equations
and associated computations }

\author{John D. Gibbon$^1$, Nairita Pal$^2$, Anupam Gupta$^3$, and Rahul Pandit$^{2}$} 
\affiliation{$^{1}$Department of Mathematics, Imperial College London, London SW7 2AZ, UK.\\
$^{2}$Centre for Condensed Matter Theory, Department of Physics, Indian Institute of Science, Bangalore, 560 012, India. \\
$^{3}$Department of Physics, University of Rome `Tor Vergata', 00133 Roma, Italy.}

\begin{abstract}
We consider the 3D Cahn-Hilliard equations coupled to, and driven by, the
forced, incompressible 3D Navier-Stokes equations. The combination, known as
the Cahn-Hilliard-Navier-Stokes (CHNS) equations, is used in statistical
mechanics to model the motion of a binary fluid. The potential development of
singularities (blow-up) in the contours of the order parameter $\phi$ is an
open problem. To address this we have proved a theorem that closely mimics the
Beale-Kato-Majda theorem for the 3D incompressible Euler equations [Beale
\textit{et al.} Commun. Math. Phys., \textbf{94} 61–66 (1984)]. By taking an
$L^{\infty}$ norm of the energy of the full binary system, designated as
$E_{\infty}$, we have shown that $\int_{0}^{t}E_{\infty}(\tau)\,d\tau$ governs
the regularity of solutions of the full 3D system. 
%$256^{3}$ and $128^{3}$ 
%numerical simulations  
Our direct numerical simulations (DNSs), of the 3D CHNS equations, for (a) a
gravity-driven Rayleigh Taylor instability and (b) a constant-energy-injection
forcing, with $128^3$ to $512^3$ collocation points and over the duration 
of our DNSs, confirm that $E_{\infty}$ remains bounded as far as 
our computations allow.
\end{abstract}

%\pacs{47.27.Ak, 52.30.Cv, 47.27.ek, 02.30.Jr}

\date{\today} 

%\keywords{Spectral Methods, particle-laden flows, phase-field model}
\keywords{Navier-Stokes equations, Integro-partial differential equations, 
Existence, uniqueness, and regularity theory, Two-phase and multiphase flows}
\pacs{47.10.A-,47.27.E-,47.27.ek,Navier-Stokes equations, Integro-partial differential 
equations, Existence, uniqueness, and regularity theory, Two-phase and multiphase flows}

\maketitle
%\ben
%\item \textbf{\color{red}No refs in text to Frisch95, Gupta \& Sbragaglia and Akshay}.
%\item \textbf{\color{blue}\S\ref{Einfsubsect} needs further remarks on the numerical results.}
%\item \textbf{\color{blue}See remarks on a change of label in equn. (10).}
%\item \textbf{\color{blue}Please note that all the aliases for $\bu$, $\bom$ etc are defined in the preamble of the source file.}
%\item \textbf{\color{blue}On reversion to ``twocolumn-style'' please keep the Appendix in widetext format by restoring 
%the commented-out the begin/end widetext-commands. }
%\item \textbf{\color{blue}I would prefer to change the $m$-label on $E_{m}$ to (say) $E_{n}$ in (\ref{Em}). We have used $2m$ for all other 
%$L^{2m}$-norms so the reader will assume the appearance of $m$ alone to be a misprint. }
%\een

\section{Introduction}\label{intro}

The Navier-Stokes (NS) equations
\cite{navier1823memoire,leray1934mouvement,batchelor1967introduction,fefferman2006existence,foias2001navier,doering1995applied},
the fundamental partial differential equations (PDEs) that govern viscous fluid
dynamics, date back to 1822.  Since its introduction in 1958 the Cahn-Hilliard
(CH) PDE \cite{cahn1958free}, the fundamental equation for the statistical
mechanics of binary mixtures, has been used extensively in studies of critical
phenomena, phase transitions
\cite{chaikin2000principles,hohenberg1977theory,lifshitz1961kinetics,gunton1983m,bray1994theory,puri2009kinetics},
nucleation \cite{lothe1962reconsiderations}, spinodal decomposition
\cite{onuki2002phase,badalassi2003computation,perlekar2014spinodal,cahn1961spinodal,berti2005turbulence},
and the late stages of phase separation
\cite{berti2005turbulence,boffetta2012two}.  If the two components of the
binary mixture are fluids, the CH and NS equations must be coupled, where the
resulting system of PDEs is usually referred to as Model
H~\cite{hohenberg1977theory} or the Cahn-Hilliard-Navier-Stokes (CHNS)
equations. 
\par\smallskip%\noindent
The increasing growth of interest in the CHNS equations arises from the elegant
way in which they allow us to follow the spatio-temporal evolution of the two
fluids in the mixture \textit{and the interfaces between them}. These interfaces 
are diffuse, so we do not have to impose boundary conditions on the moving 
boundaries between two different fluids, as in other methods for the simulation 
of multi-phase flows \cite{bray1994theory,muskat1,muskat2}. However, in addition 
to a velocity field $\bu$, we must also follow the scalar, order-parameter field 
$\phi$, which distinguishes the two phases in a binary-fluid mixture. Here,
interfacial regions are characterized by large gradients in $\phi$. The CHNS
equations have been used to model many binary-fluid systems that are of great
current interest\,: examples include studies of (a) the Rayleigh-Taylor
instability \cite{cabot2006reynolds,celani2009phase}\,; (b) turbulence-induced
suppression of the phase separation of the two components of the binary fluid
\cite{perlekar2014spinodal}\,; (c) multifractal droplet dynamics in a
turbulent, binary-fluid mixture \cite{nairita2016}\,; (d) the coalescence of
droplets \cite{shardt2013simulations}\,; and (e) lattice-Boltzmann treatments
of multi-phase flows \cite{perlekar2014spinodal,LBM}.   
\par\smallskip%\noindent
The system of Cahn-Hilliard-Navier-Stokes (CHNS) equations are written as
follows~\cite{celani2009phase,scarbolo2013turbulence,yue2004diffuse,scarbolo2011phase}\,:
\begin{eqnarray}\label{ns}
 \left(\partial_t + \bu\cdot \nabla\right)\bu &=& -\nabla p/\rho + \nu\nabla^{2}\bu - \alpha\bu - (\phi \nabla \mu) \nonumber\\ 
&&-A\bg + \bdf,\\
 \left(\partial_t + \bu\cdot \nabla\right) \phi & = &\gamma \nabla^2 {\mu}\,,
 \label{ch}
 \end{eqnarray}
where $p$ is the pressure, and $\rho$($=1$) is the constant density,
together with the incompressibility condition $\nabla \cdot \bu = 0$.  In
Eq.~\eqref{ns}, $\bu\equiv(u_x,\,u_y,\,u_z)$ is the fluid velocity and $\nu$ is
the kinematic viscosity. In the 2D case $u_z = 0$ and $\alpha$, the
air-drag-induced friction, should be included, but in 3D we set $\alpha = 0$.
$\phi(\bx,\,t)$ is the order-parameter field at the point $\bx$ and time $t$
[with $\phi(\bx,\,t)>0$ in the lighter phase and $\phi(\bx,\,t)<0$ in the
heavier phase].  The third term on the right-hand side of Eq.~\eqref{ns}
couples $\bu$ to $\phi$ via the chemical potential $\mu(\bx,\,t)$, which is
related to the the free energy $\mathcal{F}$ of the Cahn-Hilliard system as
follows\,:
\begin{eqnarray}\label{mudef}
\mu & = & \delta\mathcal{F}[\phi]/\delta \phi(\bx,\,t)\,,\\
\mathcal{F}[\phi] & = & \Lambda \I\left[\shalf|\nabla \phi|^{2} + (\phi^{2}-1)^{2}/(4\xi^{2})\right]dV\,,\label{Fdef}
\end{eqnarray}
where $\Lambda$ is the energy density with which the two phases mix in the
interfacial regime~\cite{celani2009phase}, $\xi$ sets the scale of the
interface width, $\sigma = 2(2^{\frac{1}{2}})\Lambda/3\xi$ is the surface tension,
$\gamma$ is the mobility \cite{yue2004diffuse} of the binary-fluid mixture, $A
= (\rho_2-\rho_1)/(\rho_2+\rho_1)$ is the Atwood number, and $\bg$ is the
acceleration due to gravity.
\par\smallskip%\noindent
While solutions of the CHNS equations have been shown to be regular in the
2D-case \cite{abels2009longtime,gal2010asymptotic}, with an equivalent body of
literature associated with the CH equations alone (mainly 2D) (see, e.g., Ref.
\cite{elliott1986cahn}) a critical issue for the 3D-CHNS system (\ref{ns}) -
(\ref{Fdef}) revolves around the smoothness of the contours of $\phi$ packed
together within the fluid interfaces.  The regularity of the solutions of the 3D 
Navier-Stokes (NS) equations alone is in itself a major open problem 
\cite{doering1995applied}; a coupling of the CH and the NS equations poses 
additional severe difficulties. For instance, how do we know whether a slope 
discontinuity, such as a cusp, might develop in a finite time in arbitrarily large 
spatial derivatives of $\phi$, thereby affecting the smoothness of these contours?  
Moreover, if such singularities do develop, how closely are they associated with 
the breakdown of regularity of the solutions of the 3D NS equations themselves? 
To answer such questions, we follow a strategy that is closely connected to an 
issue that once arose in studies of the incompressible 3D Euler equations\, (for 
a survey of the Euler literature see Refs.
\cite{majda2001vorticity,gibbon2008three,bardos2007euler}).  
Since the time of Leray \cite{leray1934mouvement,foias2001navier,doering1995applied} 
it has been known that the finiteness of $\I|\bom|^{2}dV$ pointwise in time controls the 
regularity of solutions of the 3D incompressible NS equations, where $\bom = \nabla \times 
\bu$ is the vorticity. There are also a variety of alternative time integral criteria, such as the 
finiteness of $\int_{0}^{T}\|\bu\|_{\infty}^{2}\,d\tau$ or $\int_{0}^{T}\left(\|\bom\|_{4}^{2}/
\|\bom\|_{2}\right)d\tau$. In addition, other conditions exist involving the pressure 
\cite{Tran2016}. In contrast, prior to 1984, it was not known what variables control the 
regularity of solutions of the 3D Euler equations. Beale, Kato, and Majda \cite{beale1984remarks} 
then proved that the time integral $\int_{0}^{T^*}\|\bom\|_{\infty}\,d\tau$ is the key object\,: 
if this integral becomes infinite at a finite time $T^{*}$, then solutions have lost regularity at $T^{*}$ 
(i.e., blow-up occurs), but there exists a global solution if, for every $T>0$, 
$\int_{0}^{T}\|\bom\|_{\infty}\,d\tau < \infty$. This result is now 
generally referred to as the BKM theorem. Its practical value is that only one simple 
integral needs to be monitored numerically. It also discounts the possibility that very 
large spatial derivatives of $\bu$ could develop a discontinuity if the integral is finite. 
\par\smallskip%\noindent
The main result of this paper is that we have shown that there exists a similar
result for the 3D-CHNS system. It can be expressed very simply and takes its
motivation from the energy $E(t)$ of the full system, which can be written as 
\bel{E2}
E(t) = \I \left[\shalf\Lambda|\nabla\phi|^{2} + \frac{\Lambda}{4\xi^{2}} 
\left(\phi^{2} - 1\right)^{2} + \shalf |\bu|^{2}\right]\,dV\,.
\ee 
Given that this can be viewed as a combination of squares of $L^{2}$-norms, it
suggests a corresponding $L^{\infty}$ version, which we call the
\textit{maximal energy}\footnote{The $L^{\infty}$-norm of a function is also
referred to as the sup- or maximum norm.}\,: 
\bel{Einf}
E_{\infty}(t) = \shalf\Lambda\|\nabla\phi\|_{\infty}^{2} + 
\frac{\Lambda}{4\xi^{2}} \left(\|\phi\|_{\infty}^{2} -1\right)^{2} 
+ \shalf \|\bu\|_{\infty}^{2}\,.
\ee 
In Sec. \ref{E-thm} we prove a theorem which says that $\int_{0}^{T^{*}}
E_{\infty}(\tau)\,d\tau$ is the key object that controls regularity of
solutions of the 3D CHNS equations exactly in the same fashion as
$\int_{0}^{T^*}\|\bom\|_{\infty}\,d\tau$ does for the 3D Euler equations
\cite{beale1984remarks}. The proof of the theorem is technically complicated,
so this is given in Appendix \ref{app1}. Our numerical calculations in that
Section (Fig. 1 (left)) suggest that $E_{\infty}$ is indeed finite. 
\par\smallskip%\noindent
In order to make a comparison with 3D Navier-Stokes results, we also calculate
the time dependence of scaled $L^{2m}$-norms of other fields, such as the fluid
vorticity ${\bom}$. The study of similar scaled norms has led to fruitful
insights into the solutions of the 3D
NS~\cite{donzis2013vorticity,gibbon2014regimes,gibbon2016high} and the 3D MHD
equations \cite{gibbon2016depletion}. We find that plots of all these norms,
versus time $t$, are ordered as a function of $m$ (curves with different values
of $m$ do not cross)\,; and, as $m \to \infty$, these curves approach a limit
curve that can be identified as the scaled $L^{\infty}$ norm. 
\par\smallskip%\noindent
The remainder of this paper is organized as follows: In Sec. \ref{nummeth} we
discuss the numerical methods that we use to study its solutions. Section
\ref{E-thm} is devoted to the statement of our $E_{\infty}$ theorem and
associated numerical results together with plots of the $L^{2m}$ norms
mentioned in the last paragraph\,. Section \ref{con} contains
concluding remarks.  In the Appendix we describe the details of the proof of
the theorem

%(B) Direct numerical simulations of
%the 2D CHNS equation, where we calculate via DNSs, several quantities like
%$D_m$, which we also calculate in the 3D case; this allows us to compare 2D and
%3D results (we have mentioned already that regularity of solutions of the 2D
%CHNS equations has been shown in earlier studies). We also explore the
%dependence of our results on (i) initial conditions, (ii) the Atwood number,
%and (iii) the spatial resolution of our DNS results.  (C) Details of our 3D
%DNSs, which we have not given in the main part of this Chapter.

%%%%%%%%%%%%%%%%%
\section{Numerical Methods}\label{nummeth}

We carry out direct numerical simulations (DNSs) of the 3D CHNS equations.  For
this we use a simulation domain that is a cubical box with sides of length
$2\pi$ and periodic boundary conditions in all three directions. We use $N^3$
collocation points, a pseudo-spectral method with a $1/2$- dealiasing rule, and
a second-order Adams-Bashforth method for time marching.  In our DNSs we use
the following two types of forcing: (a) In the first type, we use the
gravity-driven Rayleigh Taylor instability (RTI) of the interface of a heavy
fluid that is placed initially on top of a light fluid\,; this instability is
of great importance in inertial-confinement fusion
\cite{petrasso1994rayleigh,taleyarkhan2002evidence}, astrophysical phenomena
\cite{cabot2006reynolds}, and in turbulent mixing, especially in
oceanography~\cite{munk1998abyssal}\,.  (b) In the second type, we have a
forcing that yields a constant energy-injection
rate~\cite{bhatnagar2014universal}. In our RTI studies, there is a constant
gravitational field in the $\hat{z}$ direction\,; here we stop our DNS just
before plumes of the heavy or light fluid wrap around the simulation domain in
the $\hat{z}$ direction because of the periodic boundary conditions.  Most of
the DNSs of such CHNS problems, e.g., CHNS studies of the RTI, have been
motivated by experiments
\cite{waddell2001experimental,ramaprabhu2006limits,dalziel2008mixing}.  To the
best of our knowledge, no studies have investigated the growth of $L^{2m}$
norms of the quantities we have mentioned above. (For the RTI problem, some of
these norms have been studied~\cite{rao2016nonlinear} by using the DNS results
of Ref.~\cite{livescu2013numerical} for the miscible, two-fluid, incompressible
3D NS equations.) It behooves us, therefore, to initiate such DNS
investigations of $L^{2m}$ norms of fields in the 3D CHNS equations.
\par\smallskip%\noindent
The last-but-one term in Eq.~(\ref{ns}) is used in our DNSs of the RTI\,; 
in these studies we set the external force $\bdf = 0$. We also
carry out DNSs, with no gravity, but with a constant-energy-injection forcing
scheme in which 
\begin{equation}
\hat{\bdf} = P\Theta(k_f-k){\hat\bu}(\mathbf{k},\,t)
/(2E(k,\,t)), 
\label{aks_const}
\end{equation}
where $P$ is the energy-injection rate and $\Theta$ is the Heaviside function.
For simplicity, our CHNS description of binary-fluid mixtures assumes that
$\gamma$ is independent of $\phi$ and that both components of the mixtures have
the same viscosity.  We keep the diffusivity $D = \gamma\Lambda/\xi^{2}$
constant in all our DNSs. We give the parameters for our DNS runs {\tt{T1-T3}}
in Table~\ref{table1}.
\par\smallskip%\noindent
%We consider a fluid configuration in which there is a heavier fluid on top of a lighter fluid.  One major difficulty in simulating Rayleigh Taylor 
%flows is tracking the movement of the interface between the fluids, and calculating the mass, momentum and energy flux across the interface.  
%We show that using the CHNS approach, we are able to calculate the growth of $L^{2}$ norms of the various quantities in the RTI flow, in an 
%efficient manner.
%\par\smallskip%\noindent
%We define the density distribution for this two-fluid configuration in terms of $\phi$. In Fig.~\ref{def} we give a representative 
%pseudocolor plot of the time evolution of the $\phi$ field. We start with a configuration in which we have the heavy (blue) fluid 
%(with $\phi\left(\bx,\,t\right)<0$) is placed on top of the light (red) fluid (with $\phi\left(\bx,\,t\right)>0$)\,; the acceleration due 
%to gravity acts downwards.  
\begin{table}
\resizebox{0.8\linewidth}{!}
{
\begin{tabular}{|l|l|l|l|l|l|l|l|l|l|l|}
\hline
& $N$ & $A$ & $\nu$ & $D$ & $\sigma$ & ${\rm Ch}$ & ${\rm Gr}$&${\rm Re}_{\lambda}$\\
\hline
\hline
{\tt T1} &  $256$ & $0.5$ & $0.00116$ & $0.0015$ & $0.23$ & $0.011$  &&\\
{\tt T2} &  $128$ & $0$ & $0.0116$ & $0.0015$ & $0.23$ & $0.011$ & $1.2 \times 10^7$&$42.23$\\
{\tt T3} &  $512$ & $0$ & $0.00116$ & $0.0015$ & $0.23$ & $0.011$ & $1.2\times 10^9$&$300$\\
\hline
\end{tabular}
}
\caption{ The parameters $N$, $A$, $\nu$, $D$, $\sigma$, ${\rm Ch}$ and ${\rm
Gr}$ for DNS runs {\tt T1-T3}. The number of collocation points is $N^3$, $A$
is the Atwood number, $\nu$ is the kinematic viscosity, $D$ is the diffusivity, $\sigma$ is the surface
tension, ${\rm Ch}$ is the Cahn number, and ${\rm Gr}$ is the Grashof number in runs
{\tt{T2}} and {\tt{T3}}.}
\label{table1}
\end{table}
\begin{figure*}
\includegraphics[width=.5\linewidth]{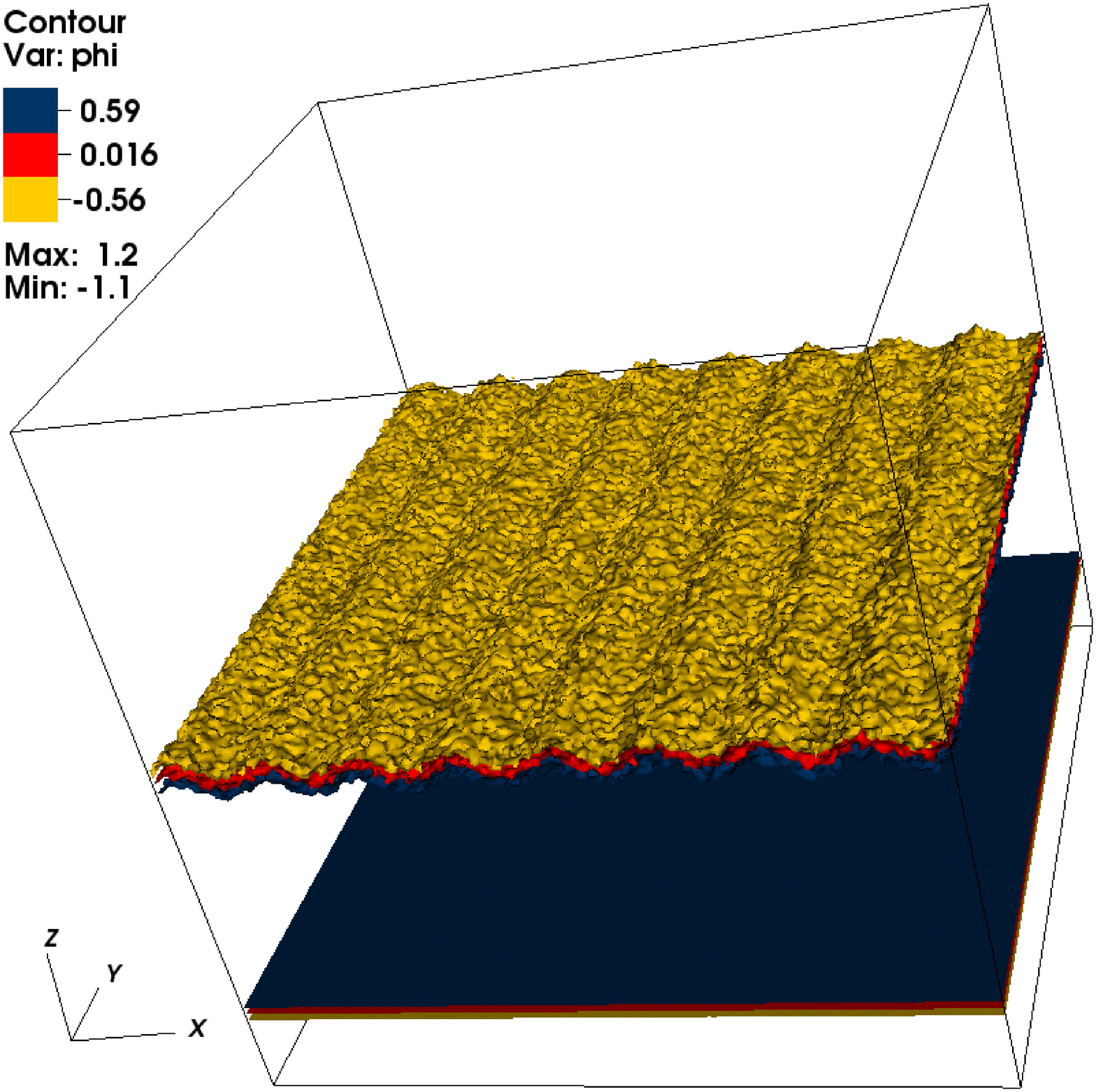}
\put(-40,195){\bf (a)}
\includegraphics[width=.5\linewidth]{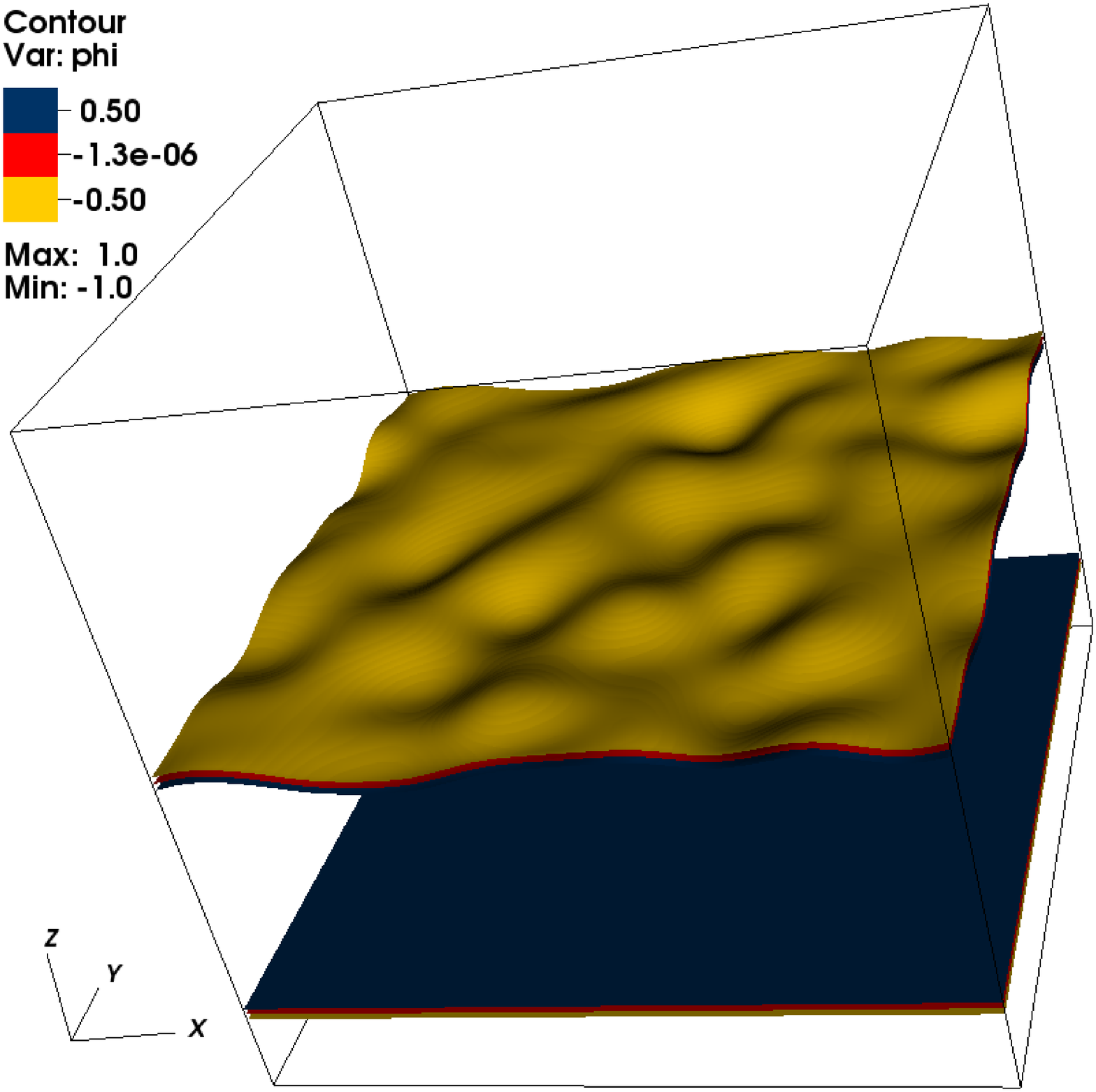}
\put(-40,195){\bf (b)}

\includegraphics[width=.5\linewidth]{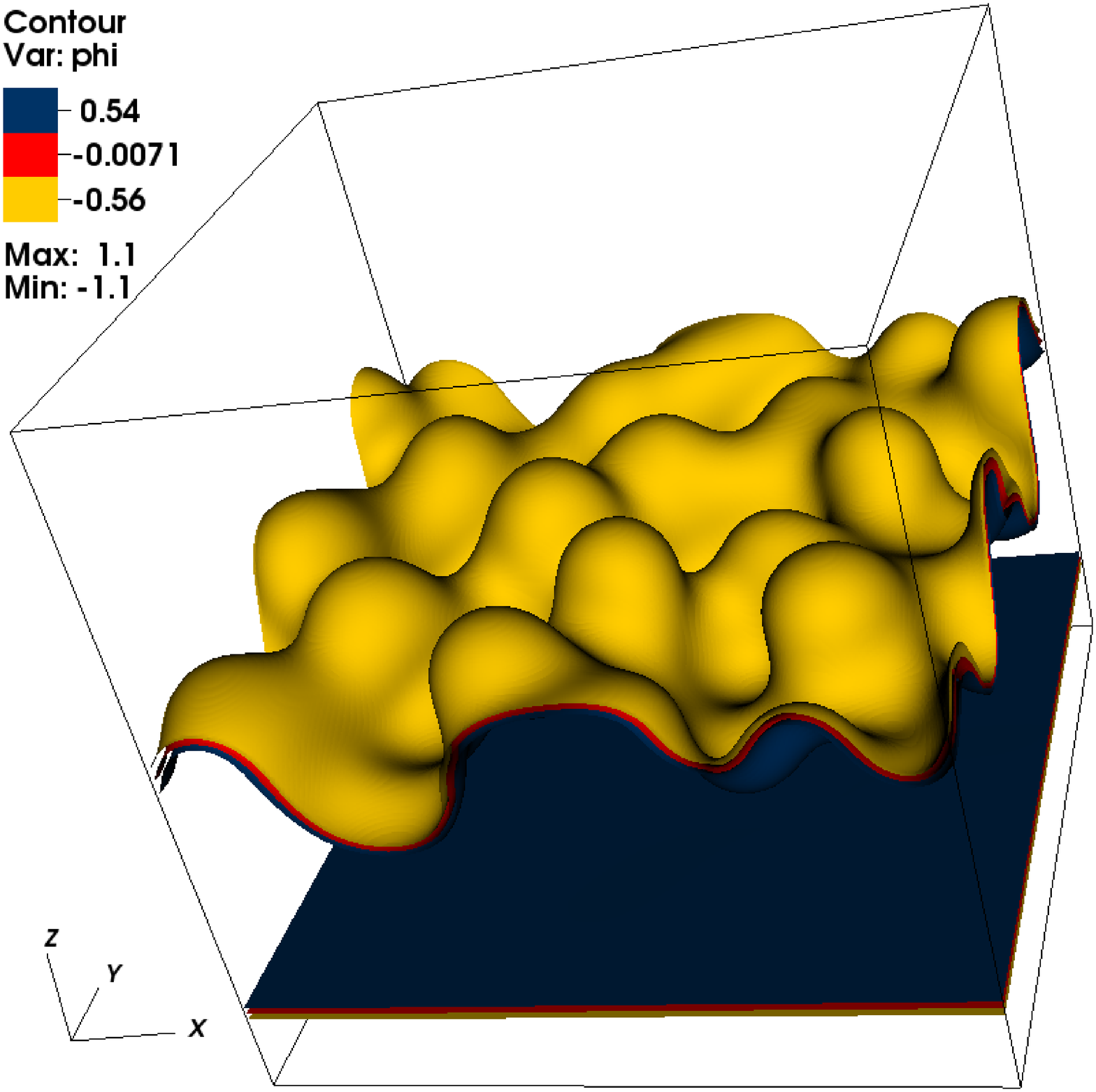}
\put(-40,195){\bf (c)}
\includegraphics[width=.5\linewidth]{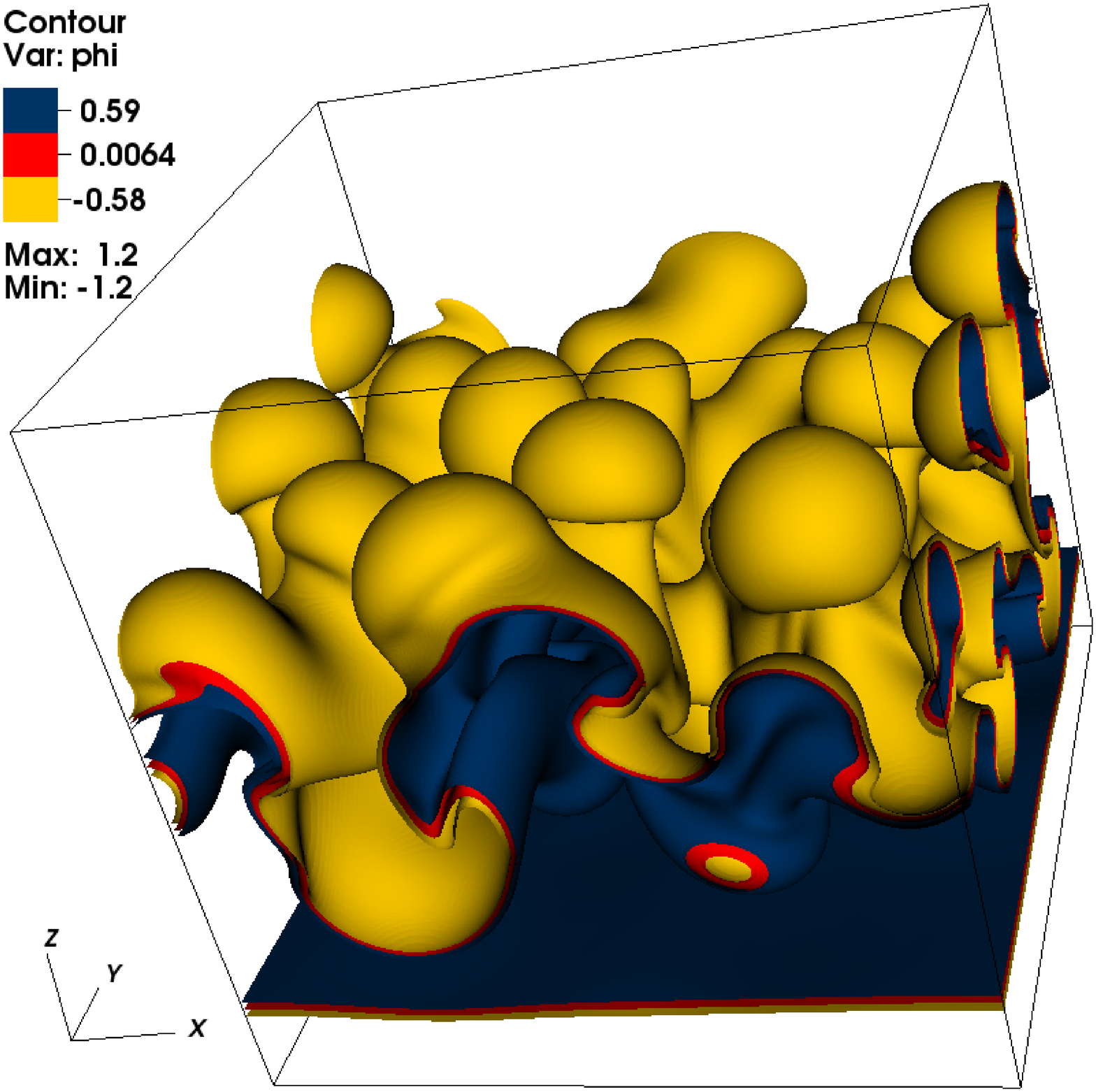}
\put(-40,195){\bf (d)}
%\includegraphics[width=.32\linewidth]{local_slope_kolmo_1024}
%\put(-20,105){\bf (c)}

%\includegraphics[width=.32\linewidth]{zoomed_local_slope_kolmo_1024}
%\put(-20,105){\bf (d)}
%\includegraphics[width=.32\linewidth]{slope_comp}
%\put(-20,105){\bf (e)}

%\includegraphics[width=.5\linewidth]{am_kolmo_newrun}
%\put(-185,160){\bf (c)}
%\includegraphics[width=.5\linewidth]{lambda_kolmo_newrun}
%\put(-185,160){\bf (d)}
\caption{(Color online) Isosurface plots of the $\phi$-field in the 3D CHNS
equations illustrating the development of the RTI with large-wavelength
perturbations in 3D (DNS run {\tt{T1}} in Table~\ref{table1}), with $256^3$
collocation points, at times (a)~$t=1$,~~(b)~$t=10$,~~(c)~$t=25$, and (d)
$t=36$.  The spatiotemporal development of this field is given in the Video {\tt{RTI\_Atwood=5e-1}}
in You Tube \cite{suppmat}.  }
\label{3D_pseudocolor}
\end{figure*}
%
%
%%%%%%%%%%%%%%%%%%%%%%%%%%%%%%%%%%%%%%%%%%%%%%
\section{$E_{\infty}$-theorem and corresponding numerical results}\label{E-thm}

In the first part of this Section (subsection A) we present our $E_{\infty}$
theorem.  We then present our numerical results in subsection B.

%%%%%%%%%%%%%%
\subsection{An $E_{\infty}$-theorem}\label{Einfsubsect}

Let us consider $n$ derivatives of both $\bu$ and $\phi$ within $L^{2}$-norms such that, for $n\geq 0$,
\bel{HPdef}
H_{n} = \I |\nabla^{n}\bu|^{2}dV\,\quad{\rm and}\quad P_{n} = \I |\nabla^{n}\phi|^{2}dV\,.
\ee
Then the CHNS equivalent of the BKM theorem \cite{beale1984remarks} is the 
following (which we prove in Appendix \ref{app1})\,: 
\begin{theorem}\label{thm1}
Consider the CHNS equations on a periodic domain $\mathcal{V} = [0,\,L]^{3}$ in three spatial dimensions. For initial data 
$u_{0} \in H_{m}$, for $m > 3/2$, and $\phi_{0}\in P_{m}$, for $m > 5/2$, suppose there exists a solution on 
the interval $[0,\,T^{*})$ where $T^{*}$ is the earliest time that the solution loses regularity, then
\bel{thm1a}
\int_{0}^{T^{*}}E_{\infty}(\tau)\,d\tau = \infty\,.
\ee
Conversely, there exists a global solution of the 3D CHNS equation if, for every $T > 0$,
\bel{thm1b}
\int_{0}^{T}E_{\infty}(\tau)\,d\tau < \infty\,.
\ee
\end{theorem}
\par
\begin{figure*}
%\put(-45,180){\bf (a)}
\includegraphics[width=0.48\linewidth]{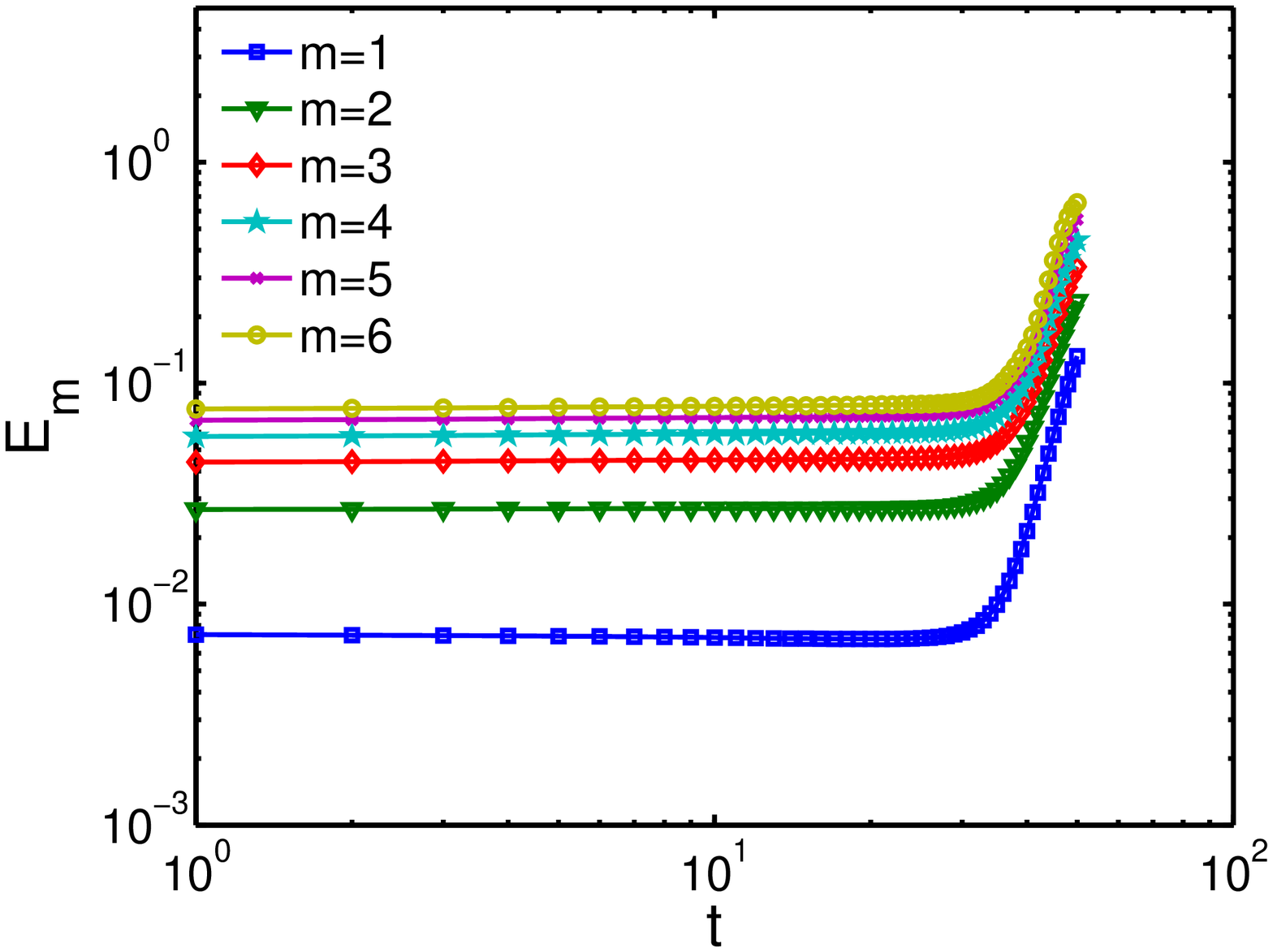}
\put(-65,45){\bf (a)}
\includegraphics[width=0.48\linewidth]{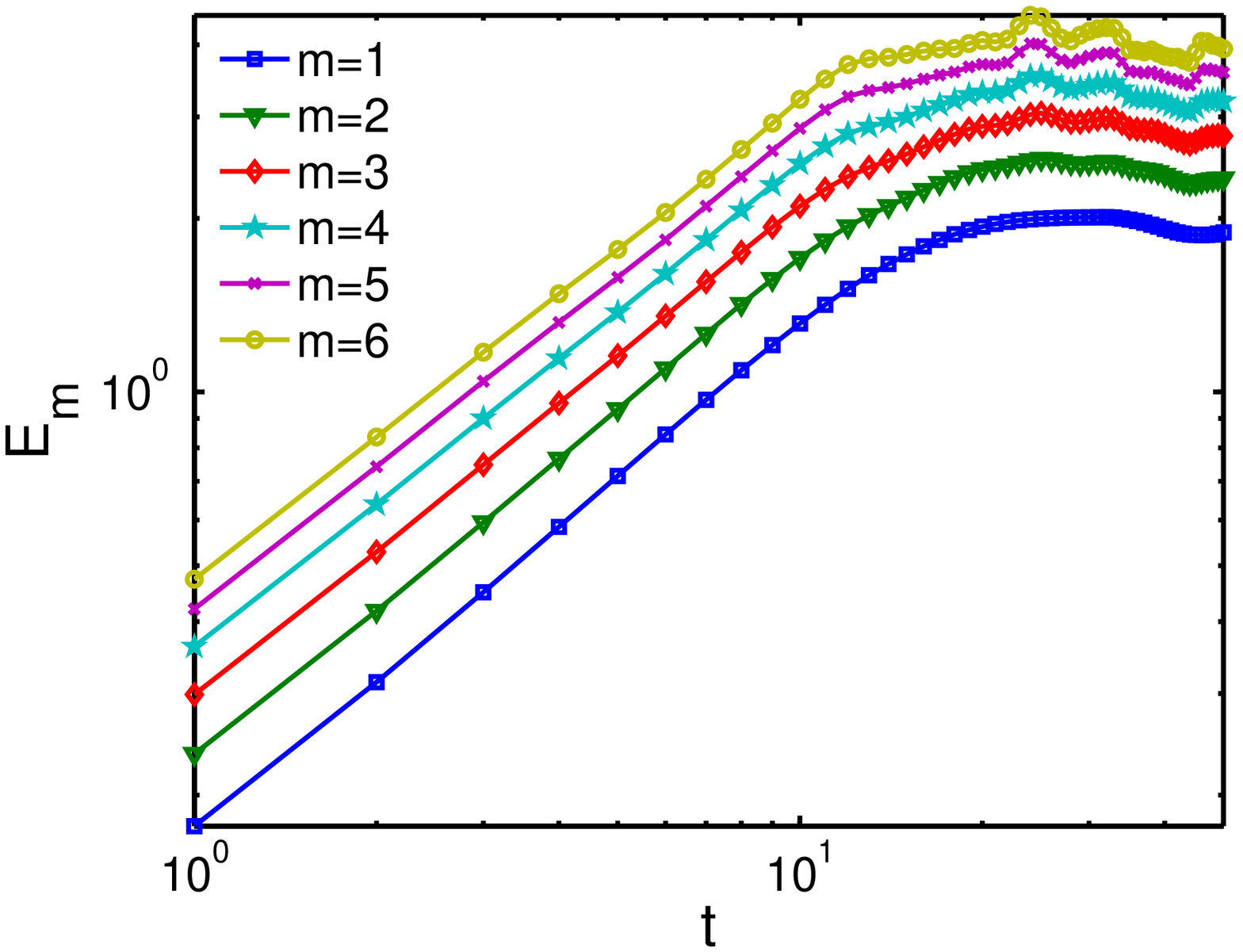}
\put(-65,45){\bf (b)}

\includegraphics[width=0.48\linewidth]{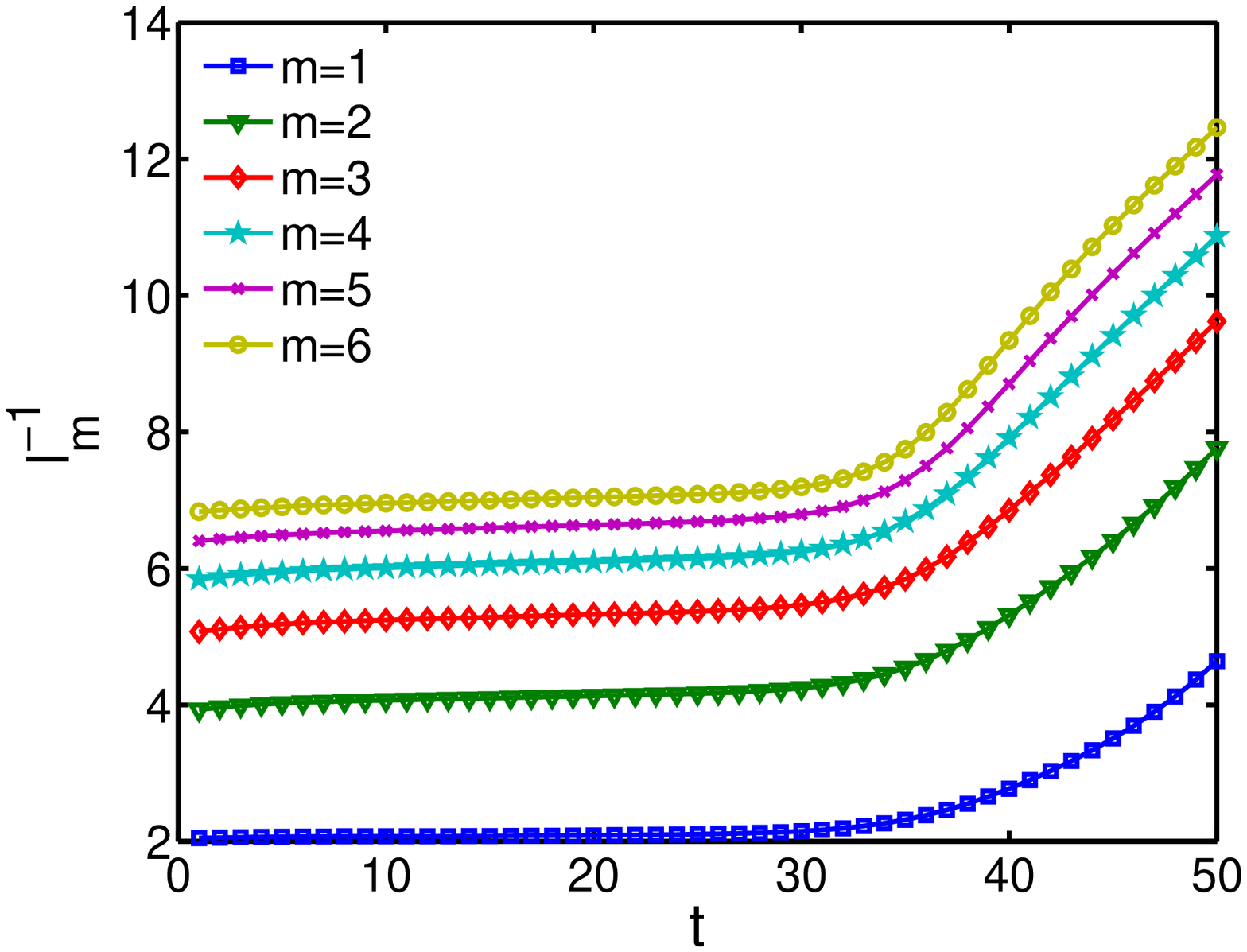}
\put(-65,155){\bf (c)}
\includegraphics[width=0.48\linewidth]{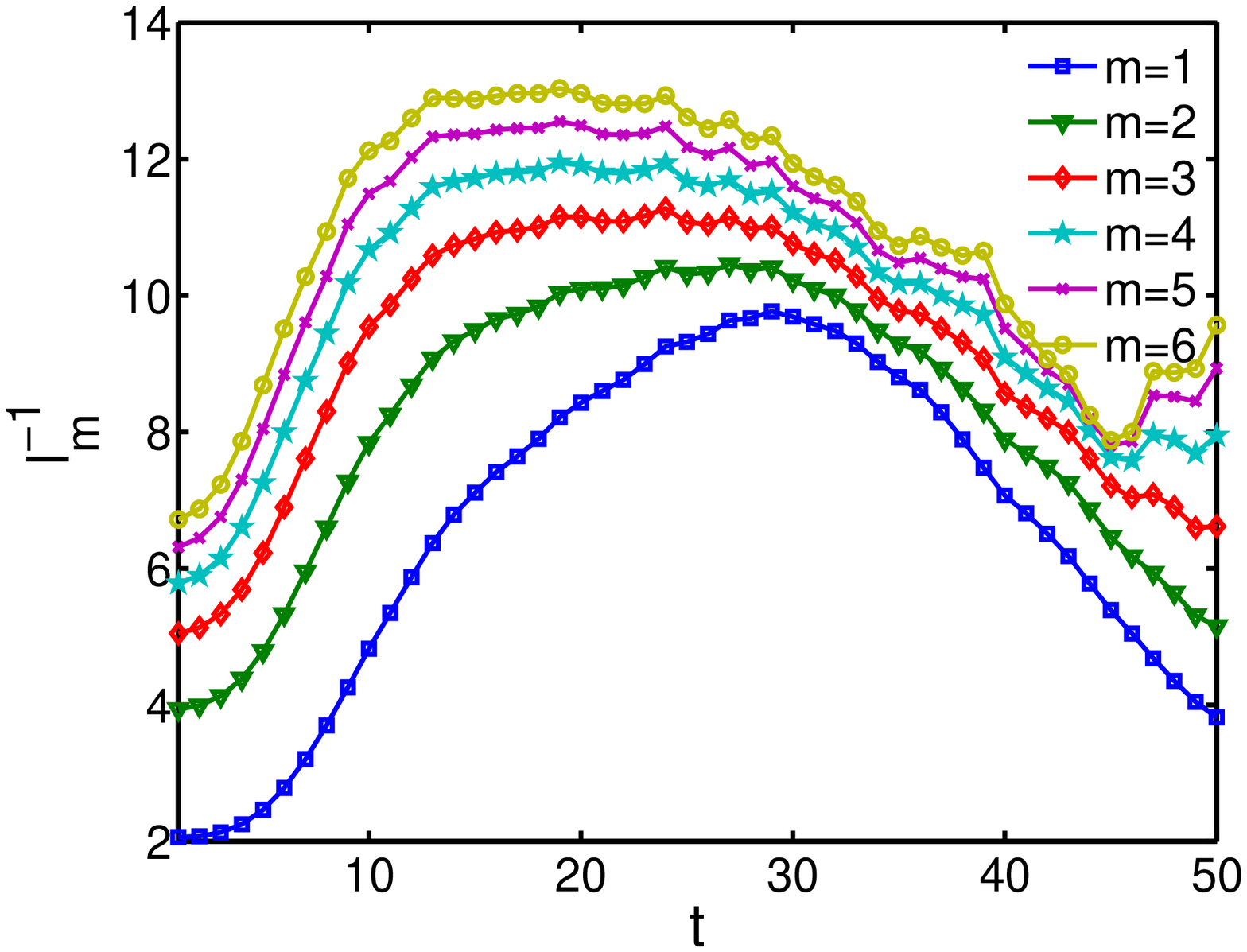}
\put(-65,155){\bf (d)}
\caption{(Color online) (a) Plots against time $t$ of $E_{m}$
according to Eq.~\eqref{Em} for 
$m=1$ (blue curve with squares), $m=2$ (green curve with inverted triangles), $m=3$ (red curve with diamonds),
$m=4$ (light blue curve with pentagrams), $m=5$ (magenta curve with crosses), and
$m=6$ (yellow curve with open circles).
These plots are for an RTI flow.
(b) Plots against time $t$ of $E_{m}$ 
according to Eq.~\eqref{Em} for 
$m=1$ (blue curve with squares), $m=2$ (green curve with inverted triangles), $m=3$ (red curve with diamonds),
$m=4$ (light blue curve with pentagrams), $m=5$ (magenta curve with crosses), and
$m=6$ (yellow curve with open circles).
These plots are for a flow with a constant-energy-injection 
forcing scheme (see Eq.~\eqref{aks_const}), with no gravity.
(c) Plots against time $t$ of $\ell_{m}^{-1}$ for $m=1$ (blue curve with squares), $m=2$ (green curve with inverted triangles), $m=3$ (red curve with diamonds),
$m=4$ (light blue curve with pentagrams), $m=5$ (magenta curve with crosses), and
$m=6$ (yellow curve with open circles). These plots are for an RTI flow.
(d) Plots against time $t$ of $\ell_{m}^{-1}$ for $m=1$ (blue curve with squares), $m=2$ (green curve with inverted triangles), $m=3$ (red curve with diamonds),
$m=4$ (light blue curve with pentagrams), $m=5$ (magenta curve with crosses), and
$m=6$ (yellow curve with open circles). These plots are for a flow with a constant-energy-injection forcing scheme. }
\label{Emlmplots}
\end{figure*}

\smallskip\noindent
The finiteness, or otherwise, of $E_{\infty}(t)$ is thus critical to the
regularity of solutions. This needs to be tested numerically from different
initial conditions. One way is to plot finite $L^{m}$-norms of the energy, namely,
\bel{Em}
E_{m}(t) = \shalf\Lambda\|\nabla\phi\|_{m}^{2} + \frac{\Lambda}{4\xi^{2}} \left(\|\phi\|_{m}^{2} - 1\right)^{2} + \shalf \|\bu\|_{m}^{2},
\ee
for increasing values of $m \geq 1$. We observe that $E_{m}(t)$ converges as
$m$ increases\,: see Fig. \ref{Emlmplots} (top panels).  This suggests that the
integral criterion within Theorem \ref{thm1} is indeed finite and thus
(\ref{thm1b}) holds leading to the regularity of these solutions, at least for
the DNSs we carry out.

%%%%%%%%%%%%%%%

%%%%%%%%%%%%%%%%%%%%%%%%%%%
\begin{figure*}

%\includegraphics[width=0.5\linewidth]{omega_3D_256}
%\put(-185,140){\bf (a)}
\includegraphics[width=0.32\linewidth]{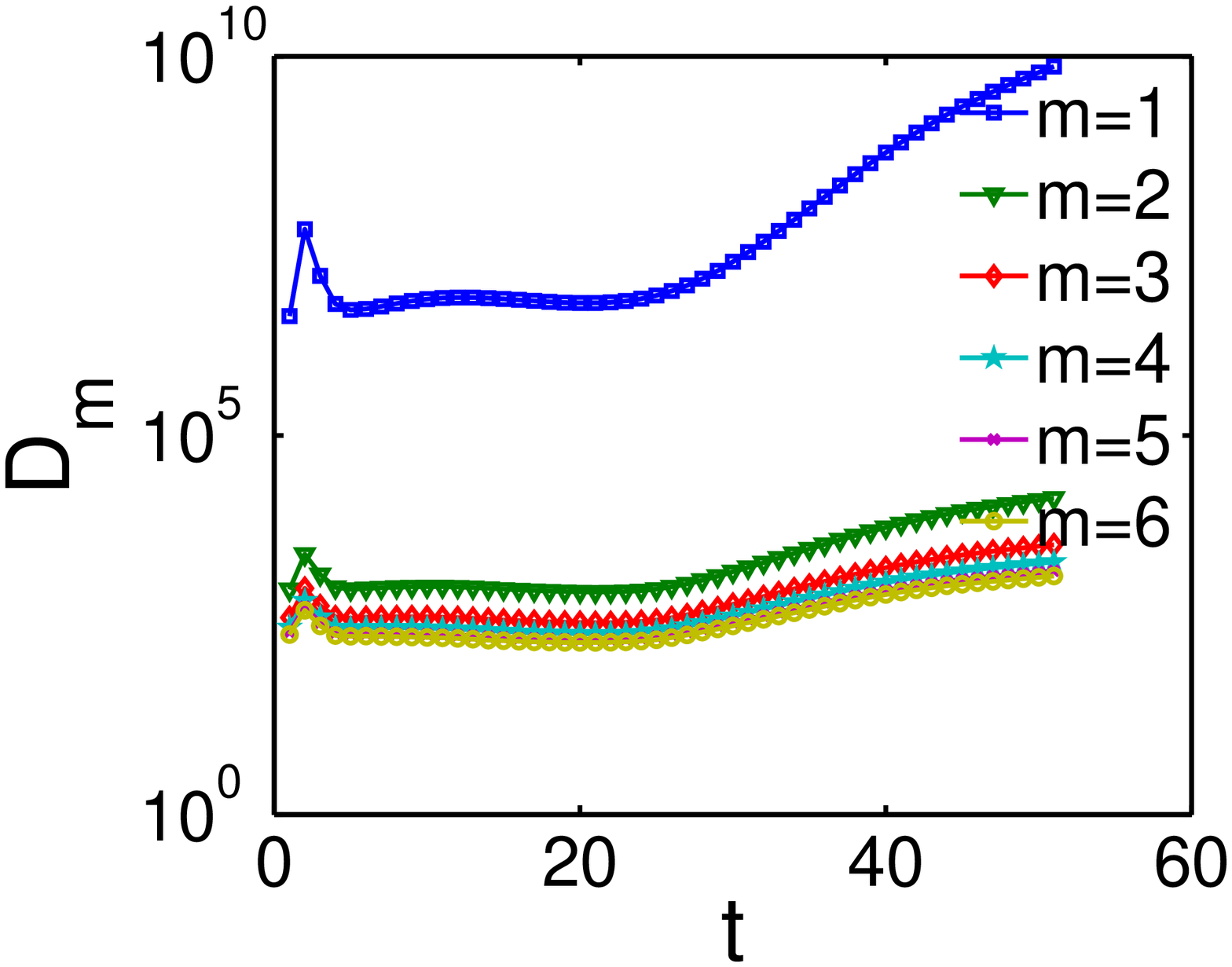}
\put(-125,100){\bf (a)}
\includegraphics[width=0.32\linewidth]{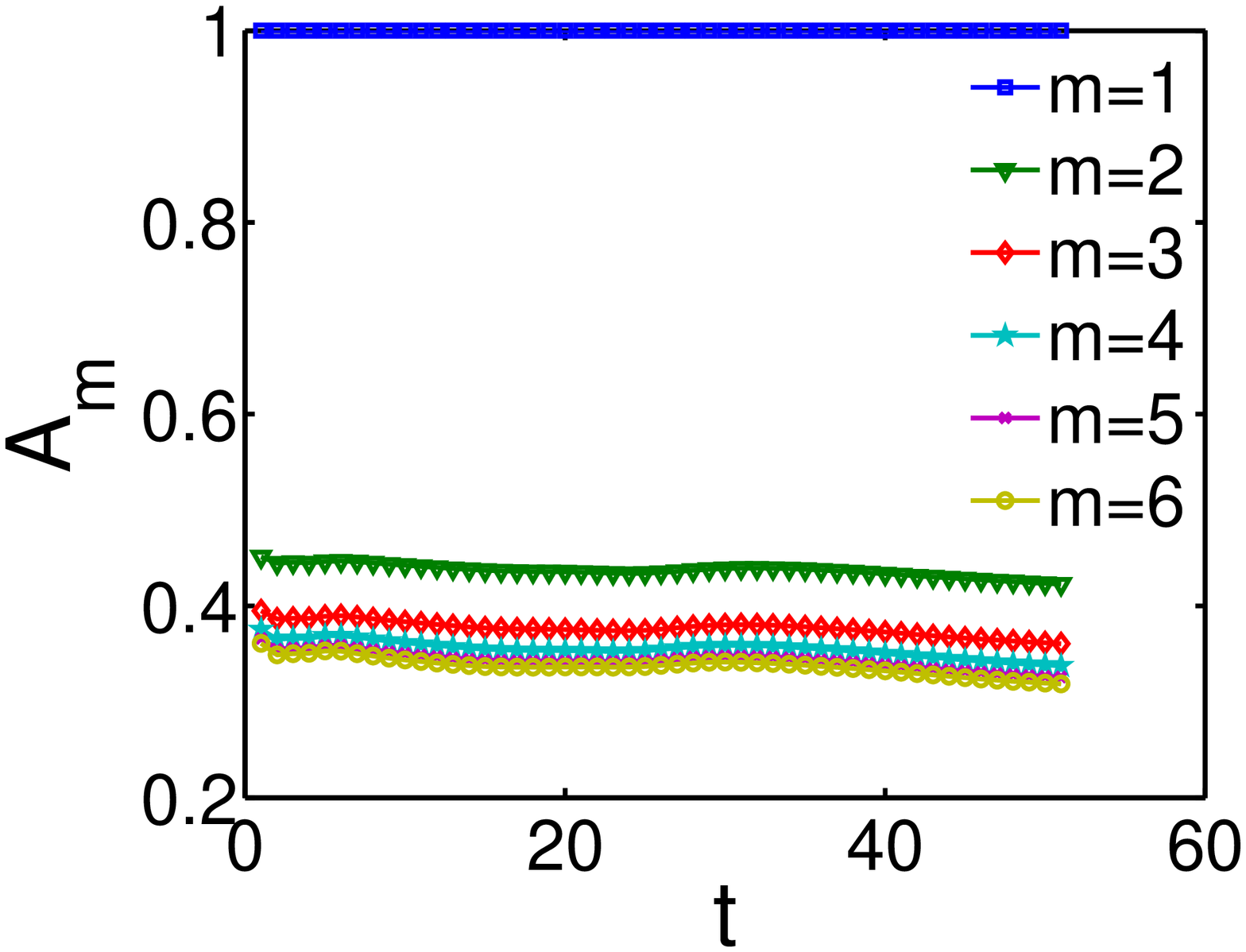}
\put(-125,100){\bf (b)}
\includegraphics[width=0.32\linewidth]{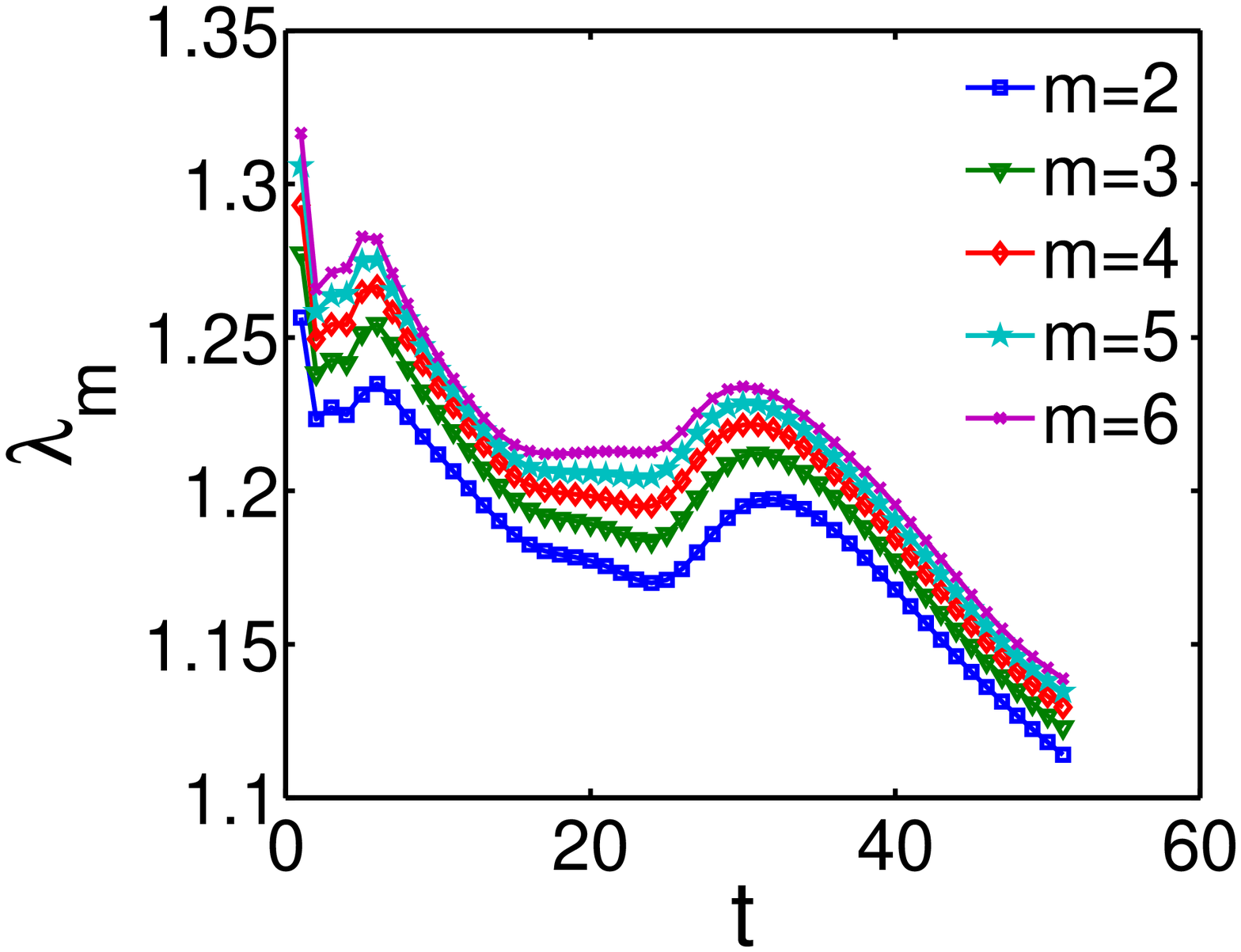}
\put(-125,100){\bf (c)}

%\includegraphics[width=0.5\linewidth]{../plots_dir/omega_3D_kol}
%\put(-185,140){\bf (a)}
\includegraphics[width=0.32\linewidth]{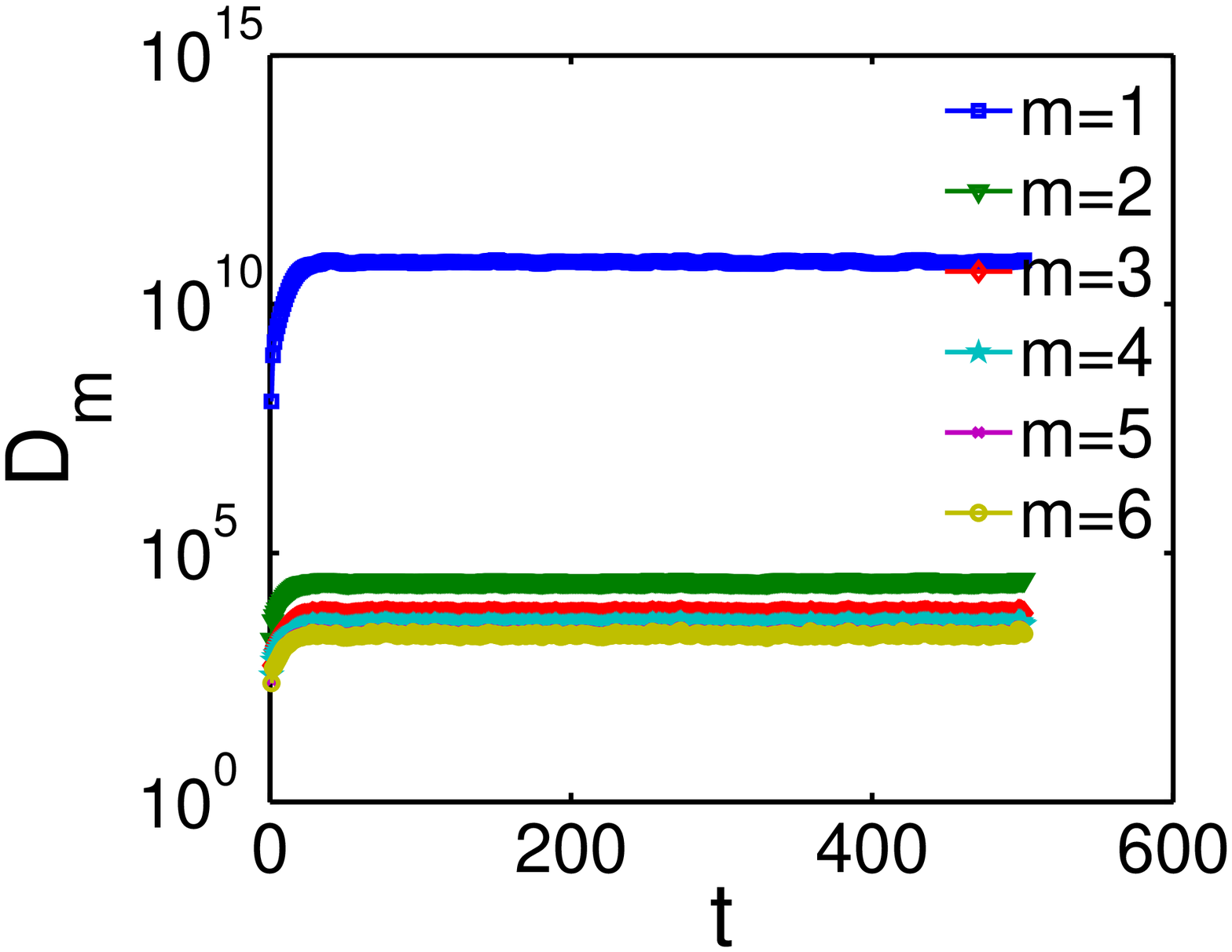}
\put(-125,100){\bf (d)}
\includegraphics[width=0.32\linewidth]{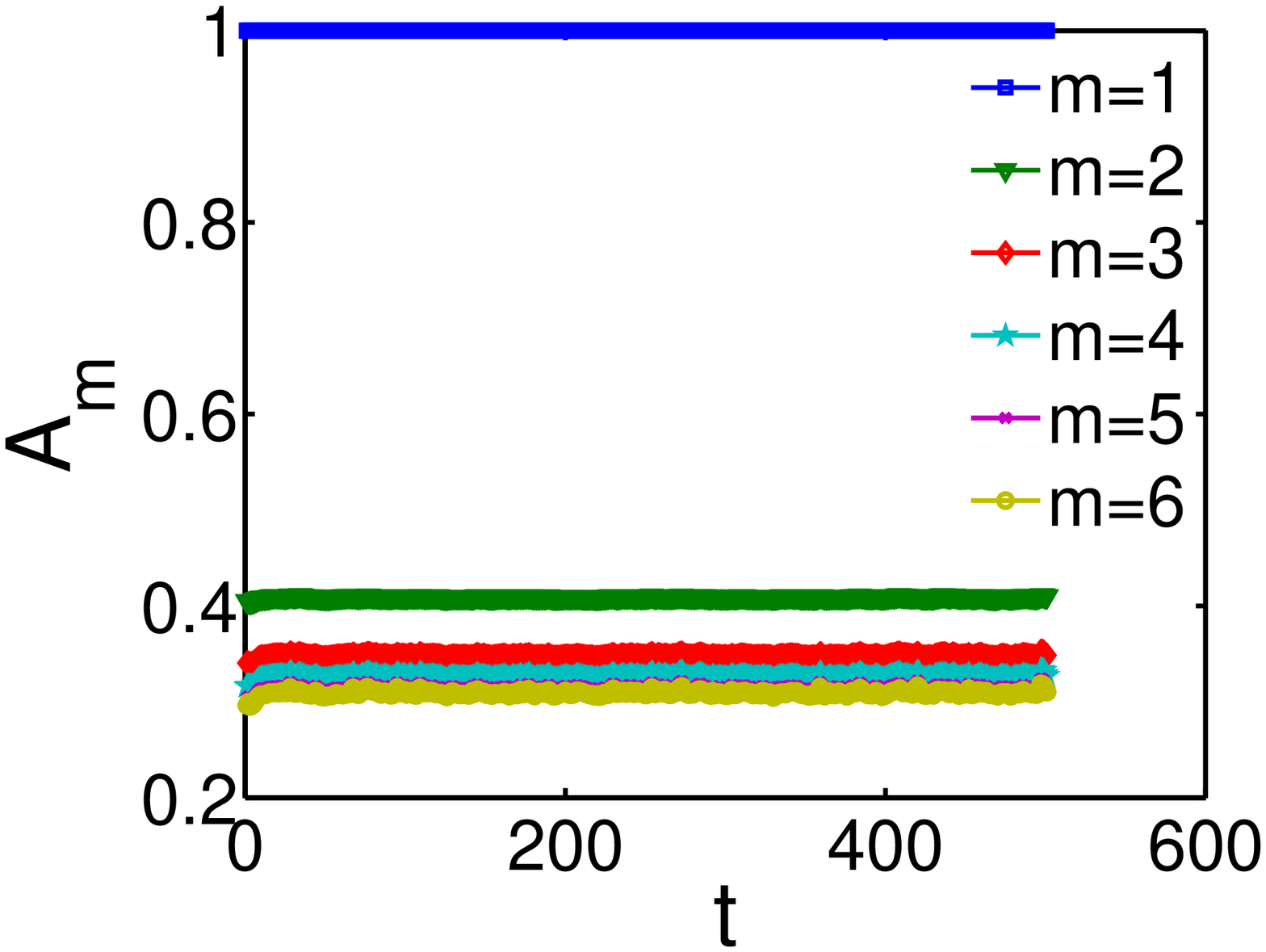}
\put(-125,100){\bf (e)}
\includegraphics[width=0.32\linewidth]{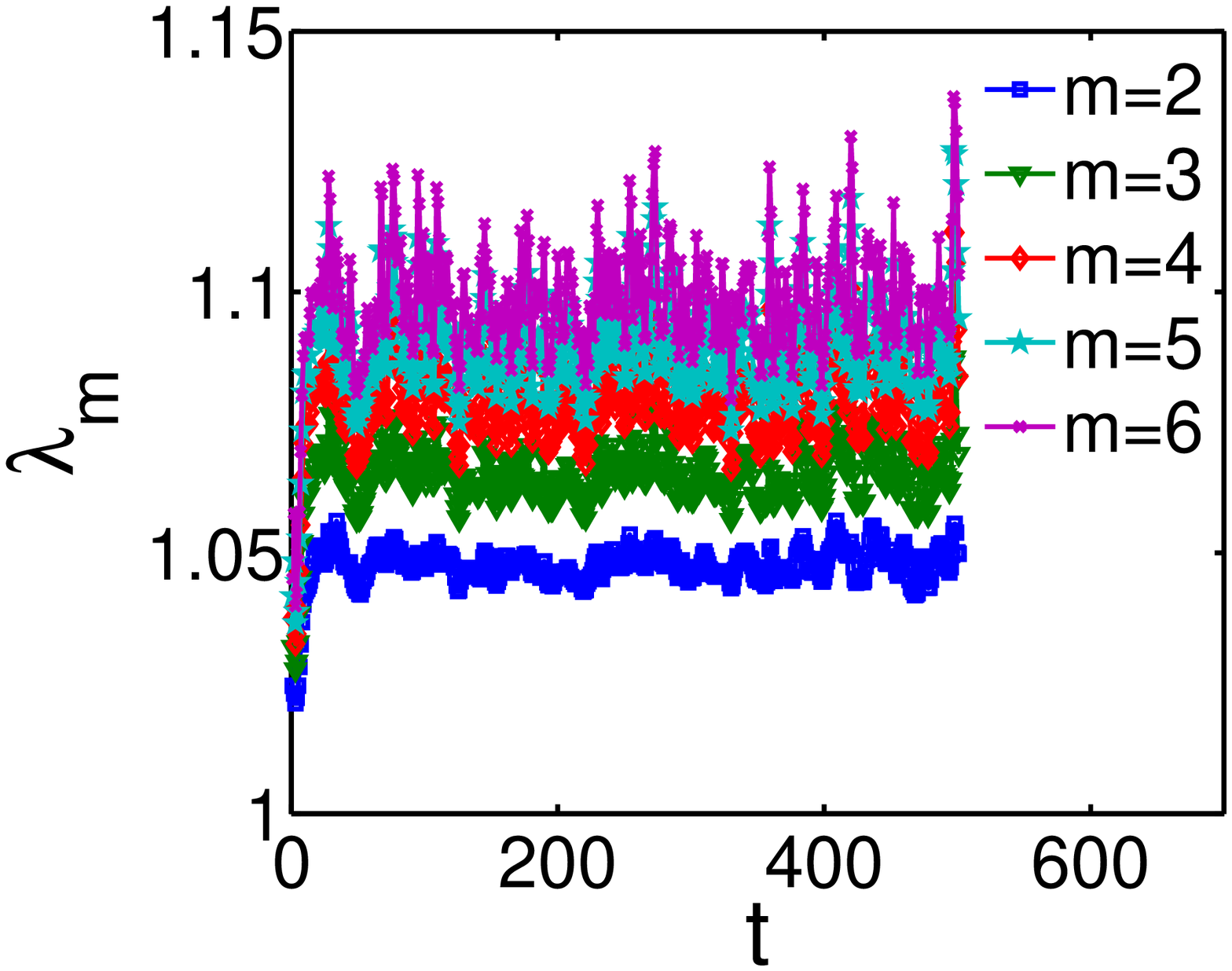}
\put(-125,100){\bf (f)}

\includegraphics[width=0.32\linewidth]{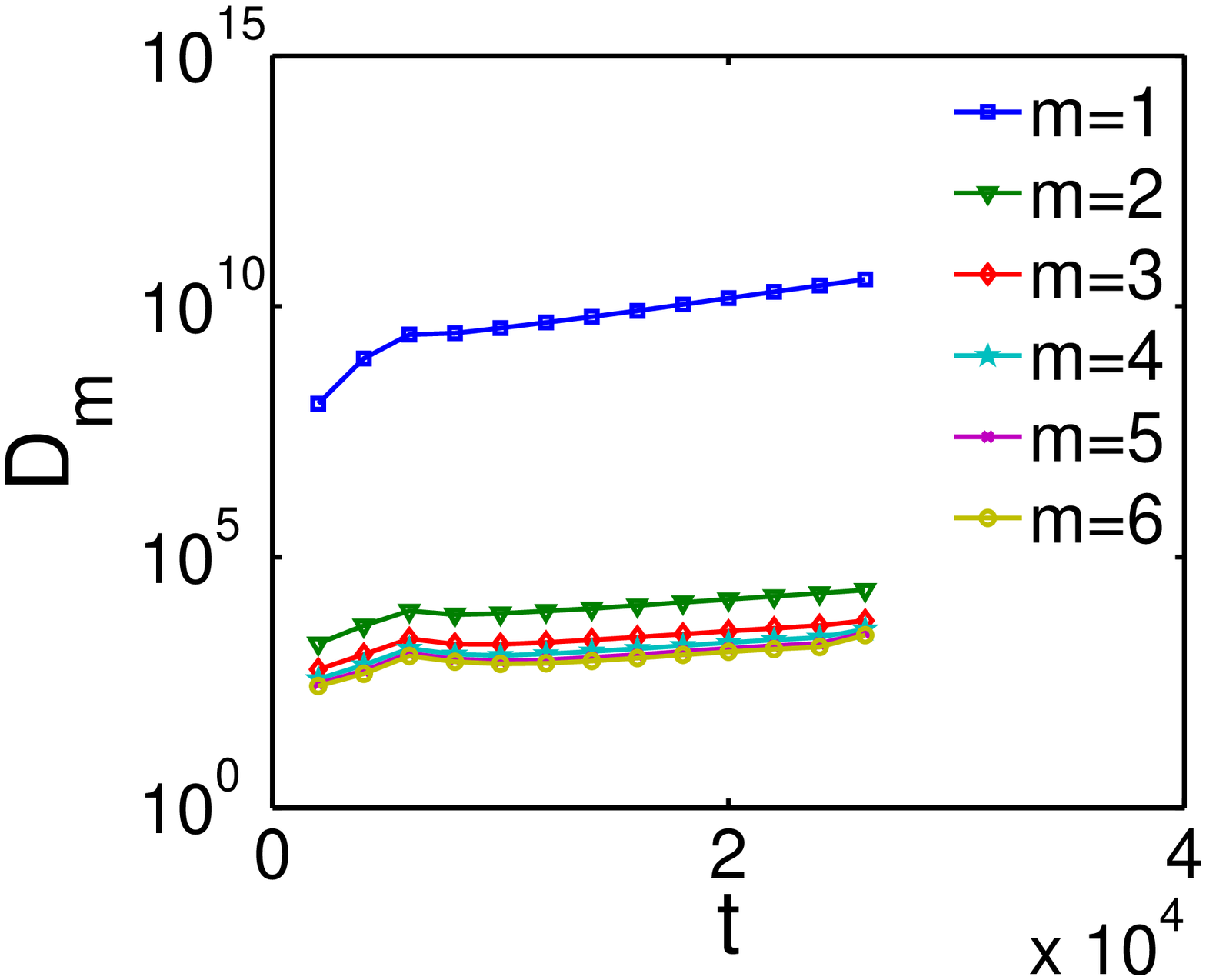}
\put(-125,100){\bf (g)}
\includegraphics[width=0.32\linewidth]{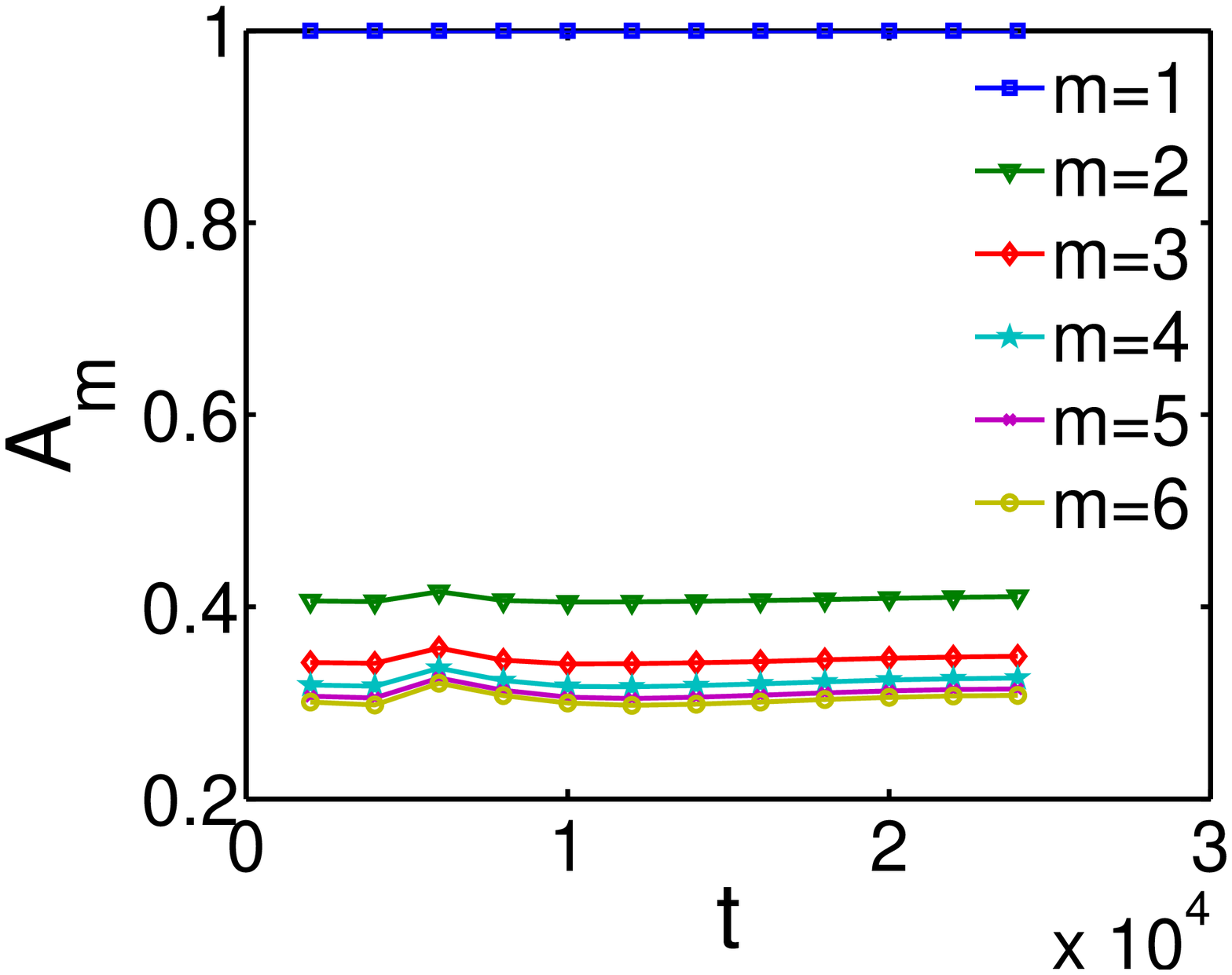}
\put(-125,100){\bf (h)}
\includegraphics[width=0.32\linewidth]{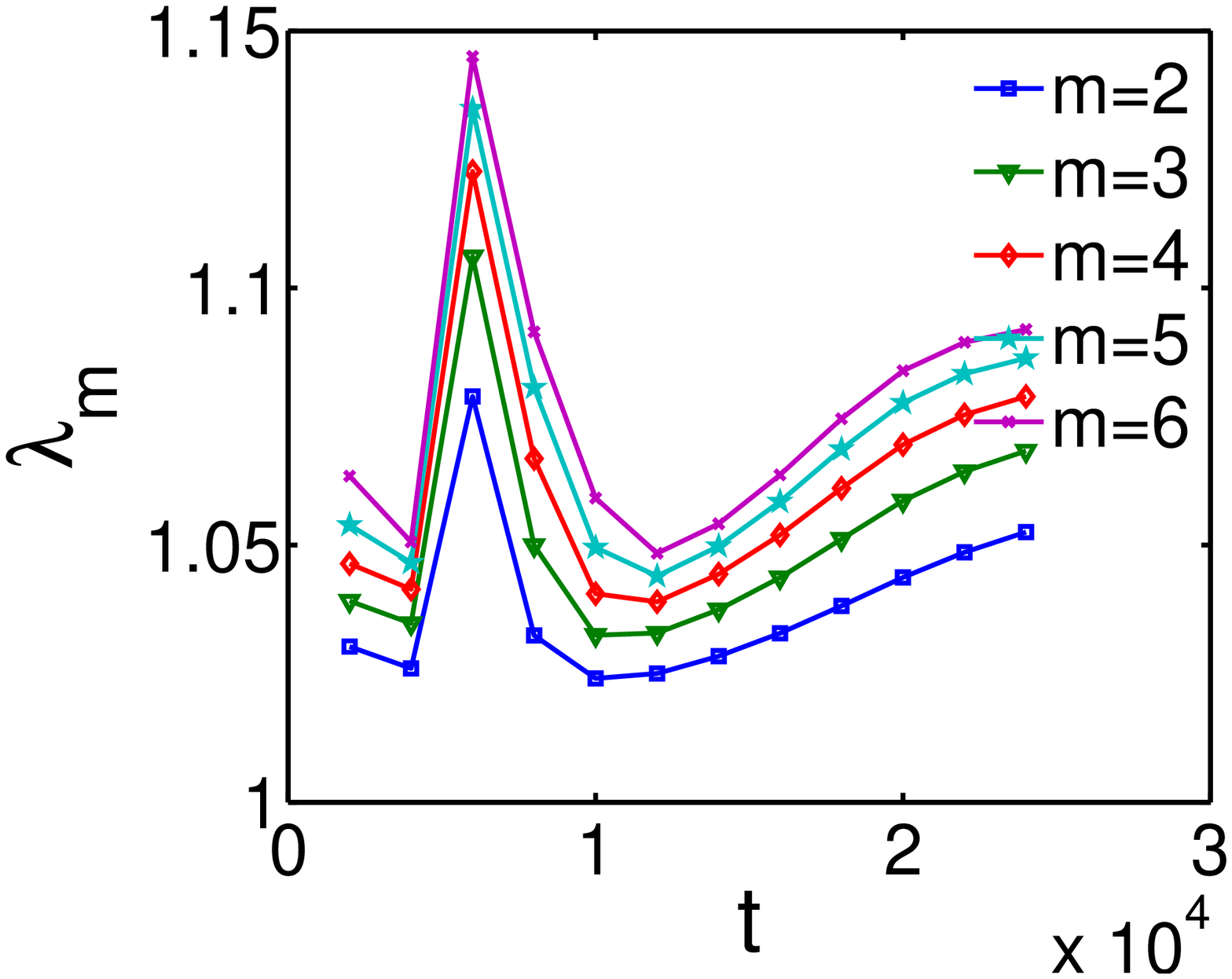}
\put(-125,100){\bf (i)}
%%%%%%%%%%%%
\caption{(Color online)  (a) Semilog (base $10$) plots versus time $t$ (with 
$256^3$ collocation (DNS run {\tt{T1}})) of (a) $D_{m}$
for $m=1$ (blue curve with squares), $m=2$ (green curve with inverted triangles), $m=3$ (red curve with diamonds), $m=4$ (light blue curve with pentagrams),
$m=5$ (magenta curve with crosses) and $m=6$ (yellow curve with open circles)\,,~~
(b) $A_{m}$ (same color conventions as in (a))\,,~~
(c) $\lambda_m(t)$ (same color conventions as in (a))\, for
the runs with RTI flows (DNS run {\tt{T1}}), and (d) $D_{m}$
for $m=1$ (blue curve with squares), $m=2$ (green curve with inverted triangles), $m=3$ (red curve with diamonds), $m=4$ (light blue curve with pentagrams),
$m=5$ (magenta curve with crosses) and $m=6$ (yellow curve with open circles)\,,~~
(e) $A_{m}$ (same color conventions as in (d))\,,~~
(f) $\lambda_m(t)$ (same color conventions as in (d))\, for
the runs with constant-energy-injection forcing
scheme (DNS run {\tt{T2}}), (g) $D_{m}$
for $m=1$ (blue curve with squares), $m=2$ (green curve with inverted triangles), $m=3$ (red curve with diamonds), $m=4$ (light blue curve with pentagrams),
$m=5$ (magenta curve with crosses) and $m=6$ (yellow curve with open circles)\,,~~
(h) $A_{m}$ (same color conventions as in (g))\,,~~
(i) $\lambda_m(t)$ (same color conventions as in (g))\, for
the runs with constant-energy-injection forcing
scheme (DNS run {\tt{T3}})
.}\label{3dplots}
\end{figure*}

%%%%%%%%%%%%%%%%%%%%%%%
\subsection{Temporal Evolution of $D_m$}\label{Dmlm}

The initial stages of the spatiotemporal development of the RTI in the 3D CHNS
system is illustrated by the isosurface plots of the $\phi$ field in
Fig.~\ref{3D_pseudocolor} (the spatiotemporal development of this field is
given in the Video {\tt{RTI\_Atwood=5e-1}} in the Supplemental Material \cite{suppmat}).  In RTI
flows, the potential energy that is stored in the initial density field is
converted to kinetic energy, which initiates fluid-mixing and a cascade of
energy from large to small length scales\,; this gives rise to filamentary
structures with enhanced gradients in $\phi$. The nonlinearity of the
binary-fluid system is responsible for this energy cascade.
%Recent studies~\cite{donzis2013vorticity,gibbon2014regimes,gibbon2016high,gibbon2016depletion}
%of solutions of the 3D NS and MHD equations have shown that the depletion of
%such nonlinearity can be investigated by examining the time evolution of the
%analogs of the $D_m$. Therefore, we begin by obtaining $D_m(t)$ for the CHNS
%equations for both RTI flows and turbulent flows driven by the type of forcing
%we have described above. 
  
For both the 3D Navier-Stokes and 3D MHD equations, a method was introduced to
estimate the degree of nonlinear depletion in the vortex stretching term(s)
\cite{donzis2013vorticity,gibbon2014regimes,gibbon2016high,gibbon2016depletion}.
It involved the use of the following $L^{2m}$-norms of the vorticity field
$\bom = \nabla \times\bu$ defined by $\left(1\leq m < \infty \right)$,
\bel{omegam}
\Omega_{m}(t) = \left( L^{-3}\I|\bom|^{2m}dV \right)^{1/2m}\,,
\ee
and also the following scaled dimensionless counterparts of $\Omega_{m}$
\bel{Dm}
D_{m}(t) = \left(\varpi_0^{-1}\Omega_m\right)^{\alpha_m},\qquad \alpha_m=\frac{2m}{4m-3}\,,
\ee
where $\varpi_{0} =\nu L^{-2}$ is the box-size frequency of the periodic box. Although the $\Omega_m$s 
must obey H\"older's inequality 
\bel{Omorder}
\Omega_{m} \leq \Omega_{m+1}\,,\qquad\mbox{for}\qquad 1\leq m < \infty\,,
\ee
no such natural ordering is enforced upon the $D_{m}$, because the $\alpha_m$ decrease with $m$ (see Eq.~\eqref{Dm}). 
We give the plots for $D_m$s for the RTI case in Fig.~\ref{3dplots}(a) and for the constant-energy-injection scheme in
Fig.~\ref{3dplots}(d).
%%%%%%%%%%%%%% cut %%%%%%%%%%%%%%%
%\rem{

%}
%%%%%%%%%%%%%% end cut %%%%%%%%%%%%%%

It was shown in Ref.~\cite{gibbon2016high} that there are good reasons why
$D_{m}$ and $D_{1}$ are such that\footnote{In \cite{gibbon2016high}
a set of multiplicative positive constants $C_{m}$ were included.}
\bel{DmD1}
D_{m} \leq D_{1}^{A_{m}(t)}\,,
\ee
with the additional relation that includes the time-dependent 
exponents $\lambda_{m}$
\bel{al}
A_{m}(t)  = \frac{\lambda_{m}(t)(m-1)+1}{4m-3}\,.
\ee
It was observed numerically \cite{gibbon2014regimes} that the maxima of the
$\lambda_{m}$ lay in the range $1.15-1.45$. For purposes of comparison between
those calculations and our RTI simulation, we plot $A_{m}(t)$ versus $t$,
in Fig.~\ref{3dplots}(b), where
\bel{Amdef}
A_{m}(t)=\ln D_{m}/\ln D_{1}\,.
\ee
We observe that the $A_{m}$ do not change significantly with $t$ but that they
depend on $m$. We also give the plot of $A_m(t)$ versus $t$
for the case of constant-energy-injection in Fig.~\ref{3dplots}(e).
%In a similar manner to the 3D Navier-Stokes 
%simulations in \cite{Gibbon14}, 
As in DNSs of the 3D Navier-Stokes equation \cite{gibbon2014regimes}, we find,
for the 3D CHNS system, that $D_1$ lies well above the other $D_m$ (see
Fig.~\ref{3dplots}(a) and Fig.~\ref{3dplots}(c)). We give the plots for $\lambda_{m}(t)$ in
Fig.~\ref{3dplots}(c) (for the RTI case)
and in Fig.~\ref{3dplots}(f) (for the constant-energy-injection 
forcing scheme). In the 3D NS case, the $\lambda_{m}$ are related to the
spectral exponents for the inertial-range, power-law form of the energy spectra
\cite{gibbon2014regimes}\,; the analogous relation for the 3D CHNS case is not
straightforward because the power-law ranges in such spectra depend on several
parameters in the CHNS equations (see, e.g., Ref.~\cite{nairita2016}).
\par\smallskip%\noindent
We also compute the temporal evolution of the $L^{2m}$-norms of the gradients
of $\phi$ by using the definition of the inverse length scale $\ell_{m}^{-1}$ 
\bel{ell2def}
\ell_{m}^{-2m} = \frac{\I |\nabla\phi|^{2m}dV}{\I |\phi|^{2m}dV}\,.
\ee
Figures~\ref{Emlmplots}(a)-(d) show plots of $E_m$ and $\ell_{m}^{-1}$ versus
time $t$ for different values of $m$.  These are qualitatively similar to those
for $D_m$ in so far as curves for different values of $m$ do not cross\,; they
are ordered in $m$ such that $\ell_{m}^{-1} < \ell_{m+1}^{-1}$.  Furthermore,
both $\ell_{m}^{-1}$ and $E_m$ approach limiting curves as $m \rightarrow
\infty$\,.
% we have checked that, as $m \rightarrow \infty$, $E_m \rightarrow
%E_\infty$. 
(We mention in passing that errors increase, as $m$ increases.  We
present data for values of $m$ for which we have reliable data). This
clustering of the $E_m$ suggests convergence to a finite value of $E_{\infty}$.
 
%%%%%%%%%%%%%%%%%%%
\section{Conclusion}\label{con}

The regularity of solutions of the 3D Navier-Stokes (3DNS) equations presents
formidable difficulties.  It remains to this day one of the outstanding open
problems in modern applied mathematics \cite{fefferman2006existence}.  Coupling
these 3DNS equations to the 3D Cahn-Hilliard equations creates a set of 3D CHNS
PDEs that govern an incompressible binary fluid, but, in so doing, creates a
system where the already formidable difficulties with the 3D NS system are
amplified many times over. The elegant and powerful proofs of regularity by
Abels \cite{abels2009longtime} and Gal and Graselli \cite{gal2010asymptotic} in
the 2D case show how much harder the coupled 2D CHNS system is to deal with
than the 2D Navier-Stokes equations alone.

The main challenges in the 3D system considered here lie in the
behavior of not only arbitrarily large gradients of the velocity field $\bu$
but also of arbitrarily large gradients of $\phi$, the order parameter. The
$E_{\infty}$ theorem, stated in Sec. \ref{E-thm} and proved in Appendix
\ref{app1},  is a conditional-regularity criterion on periodic boundary
conditions that is realistically computable. The motivation for this result
lies in the BKM theorem for the 3D Euler equations. Constantin, Fefferman and
Majda \cite{constantin1996geometric} have reduced the $\|\bom\|_{\infty}$
within the BKM criterion to $\|\bom\|_{p}$ for finite $p\geq 2$, but at the
heavy price of introducing technically complicated, local constraints on the
direction of vorticity, which are difficult to compute. Thus, the original form
of the BKM theorem, with its single requirement of $\|\bom\|_{\infty}$ being
finite, remains the simplest regularity criterion to this day. Our $E_{\infty}$
theorem is the equivalent result for the 3D CHNS system.

Our curves for $E_{m}$ versus time in Fig. \ref{Emlmplots} (left) suggest
convergence to $E_{\infty}$, with increasing values of $m$, thereby indicating
that solutions remain regular for as long as our DNSs remain valid, even
though more resolution would be desirable in the future to investigate
the delicate issue of possible finite-time singularities in solutions
of the 3D CHNS equations. 

\begin{acknowledgements}

We thank  Prasad Perlekar, Samriddhi Sankar Ray, Akshay Bhatnagar, and Akhilesh
Kumar Verma for discussions.  J.D.G. thanks F\'ed\'eration Doeblin for support
and the International Centre for Theoretical Sciences, Bangalore for
hospitality during a visit in which this study was initiated.  N.P. and R.P.
thank University Grants Commission (India), Department of Science and
Technology (India), and Council of Scientific and Industrial Research (India),
 for support and SERC (IISc) for
computational resources.  
%A.G. acknowledges support from a grant from the
%European Research Council (ERC) under the European Community’s Seventh
%Framework Programme (FP7/2007-2013)/ERC Grant Agreement No. 297004.

\end{acknowledgements}

%%%%%%%%%%%%%%%%%%%
\begin{widetext}
\appendix
\section{Proof of Theorem \ref{thm1}}\label{app1}

In the following proof the coefficients in Eqs.~\eqref{ns}-\eqref{Fdef} in the main paper are set to unity to avoid needless 
complication. First, we recall the definitions of $H_{n}$ and $P_{n}$ in Eq.~\eqref{HPdef}. In addition to these we 
define 
\bel{Xndef}
X_{n} = H_{n} + P_{n+1}\,.
\ee
The proof uses the method of BKM \cite{beale1984remarks}, which is by contradiction. The strategy is the following\,: 
suppose there exists an interval $[0,\,T^{*})$ on which solutions are globally regular with the earliest loss of regularity 
at $T^{*}$. Assume that $\int_{0}^{T^{*}} E_{\infty}(\tau)\,d\tau < \infty$, and then show that a consequence of this is that 
$X_{n}(T^{*}) < \infty$, which contradicts the statement that solutions first lose regularity at $T^{*}$. This falsifies the 
assumption of the finiteness of the integral.  We proceed in 3 steps. 
\par\medskip\noindent
\textbf{Step 1\,:} We begin with the time evolution of $P_{n}$\, (the dot above $P_n$ denotes a time derivative):
\bel{pn1}
\shalf\dot{P}_{n} =  - P_{n+2} + P_{n+1} + \I (\nabla^{n}\phi)\nabla^{n}\Delta(\phi^{3})\,dV
- \I (\nabla^{n}\phi )\nabla^{n}(\bu\cdot\nabla\phi)\,dV\,;
\ee
and then we estimate the third term on the right as
\bel{pn2}
\left|\I (\nabla^{n}\phi)\nabla^{n}\Delta(\phi^{3})\,dV\right| \leq
\|\nabla^{n}\phi\|_{2}\sum_{i,j=0}^{n+2}C^{n+2}_{i,j}\|\nabla^{i}\phi\|_{p}
|\nabla^{j}\phi\|_{q}\|\nabla^{n+2-i-j}\phi\|_{r},
\ee
where $1/p + 1/q + 1/r = 1/2$. Now we use a sequence of Gagliardo-Nirenberg inequalities 

\beq{pn3}
\|\nabla^{i}\phi\|_{p} &\leq& c_{n,i} \|\nabla^{n+2}\phi\|_{2}^{a_{1}}\|\phi\|_{\infty}^{1-a_{1}},\non\\
\|\nabla^{j}\phi\|_{q} &\leq& c_{n,j} \|\nabla^{n+2}\phi\|_{2}^{a_{2}}\|\phi\|_{\infty}^{1-a_{2}},\\
\|\nabla^{n}\phi\|_{r} &\leq& c_{n,i,j} \|\nabla^{n+2-i-j}\phi\|_{2}^{a_{3}}\|\phi\|_{\infty}^{1-a_{3}}\,,\non
\eeq
where, in $d$ dimensions,
\beq{pn4}
\frac{1}{p} &=& \frac{i}{d} + a_{1}\left(\frac{1}{2} - \frac{n+2}{d}\right),\non\\
\frac{1}{q} &=& \frac{j}{d} + a_{2}\left(\frac{1}{2} - \frac{n+2}{d}\right),\\
\frac{1}{r} &=& \frac{n+2-i-j}{d} + a_{3}\left(\frac{1}{2} - \frac{n+2}{d}\right)\,.\non
\eeq

By summing these and using $1/p + 1/q + 1/r = 1/2$, it is seen that $a_{1}+a_{2}+a_{3}=1$. Thus, we have 
\bel{pn5}
\left|\I (\nabla^{n}\phi)\nabla^{n+2}(\phi^{3})\,dV\right| \leq
c_{n}\|\nabla^{n}\phi\|_{2}\|\nabla^{n+2}\phi\|_{2}\|\phi\|_{\infty}^{2}
\leq \shalf P_{n+2} + c_{n} P_{n}\|\phi\|_{\infty}^{4}\,,
\ee
and so Eq.~\eqref{pn1} becomes (here and henceforth coefficients such as $c_n$
are multiplicative constants),
\bel{pn6}
\shalf\dot{P}_{n}  =  -\shalf P_{n+2} + P_{n+1} + c_{n}\|\phi\|_{\infty}^{4}P_{n} + 
\left|\I (\nabla^{n}\phi)\nabla^{n}(\bu\cdot\nabla\phi)\,dV\right|\,.
\ee
Estimating the last term in Eq.~\eqref{pn6} we have
\beq{pn7}
\left|\I (\nabla^{n}\phi) \nabla^{n}(\bu\cdot\nabla\phi)\,dV\right| &=& 
\left|-\I (\nabla^{n+1}\phi) \nabla^{n-1}(\bu\cdot\nabla\phi)\,dV\right|\non\\
 &\leq& \|\nabla^{n+1}\phi\|_{2} \sum_{i=0}^{n-1}C^{n}_{i}\|\nabla^{i}\bu\|_{p} \|\nabla^{n-1-i}(\nabla\phi)\|_{q},
\eeq
where $1/p + 1/q = 1/2$. Now we use two Gagliardo-Nirenberg inequalities in $d$ dimensions to obtain
\beq{pn8}
\|\nabla^{i}\bu\|_{p} &\leq& c\,\|\nabla^{n-1}\bu\|_{2}^{a} \|\bu\|_{\infty}^{1-a},\\
\|\nabla^{n-1-i}(\nabla\phi)\|_{q} & \leq & c\,\|\nabla^{n-1}(\nabla\phi)\|_{2}^{b} \|\nabla\phi\|_{\infty}^{1-b}.
\label{pn8b}
\eeq
Equations~\eqref{pn8} and \eqref{pn8b} follow from
\beq{lb1}
\frac{1}{p} & = & \frac{i}{d} + a\left(\frac{1}{2} - \frac{n-1}{d}\right),\\
\frac{1}{q} & =  & \frac{n-1-i}{d} + b\left(\frac{1}{2} - \frac{n-1}{d}\right).
\eeq
Because $1/p + 1/q = 1/2$ then $a+b=1$. Thus Eq.~\eqref{pn2} turns into 
\beq{pn9}
\left|\I (\nabla^{n}\phi)\nabla^{n}(\bu\cdot\nabla\phi)\,dV\right| &\leq&
c_{n}P_{n+1}^{1/2} H_{n-1}^{a/2}P_{n}^{b/2}\|\bu\|_{\infty}^{1-a}\|\nabla\phi\|_{\infty}^{1-b}\non\\
&\leq& P_{n+1}^{1/2} \left[c_{n}H_{n-1}\|\nabla\phi\|_{\infty}^{2}\right]^{a/2}\left[P_{n}\|\bu\|_{\infty}^{2}\right]^{b/2}\non\\
&\leq& \shalf P_{n+1} + \shalf ac_{n}H_{n-1}\|\nabla\phi\|_{\infty}^{2}+ \shalf b P_{n}\|\bu\|_{\infty}^{2}\,,
\eeq
and Eq.~\eqref{pn6} becomes
\bel{pn10}
\shalf\dot{P}_{n}  =  - \shalf P_{n+2} + \threehalves P_{n+1} 
+ c_{n,1}\left(\shalf\|\phi\|_{\infty}^{4} + \|\bu\|_{\infty}^{2}\right)P_{n} 
+ c_{n,2}H_{n-1}\|\nabla\phi\|_{\infty}^{2}\,.
\ee
\par\medskip\noindent
\textbf{Step 2\,:} Now we look at $H_{n}$ defined in Eq.~\eqref{HPdef} using Eq.~\eqref{h13d} with $\bdf = - \hat{z}\phi$. The 
easiest way is to use the 3D NS equation in the vorticity form as in Doering and Gibbon \cite{doering1995applied} to obtain the 
$\|\bu\|_{\infty}^{2}$-term in Eq.~\eqref{hn1}, where gradient terms have been absorbed into the pressure term, which 
disappears under the curl-operation\,:
\bel{h13d}
\left(\partial_{t}+\bu\cdot\nabla\right)\bom = \Delta\bom + \bom\cdot\nabla\bu + \nabla\phi\times\nabla\Delta\phi 
- \nabla^{\perp}\phi\,.
\ee
Therefore,
\beq{hn1}
\shalf\dot{H}_{n} &\leq& -\shalf H_{n+1} + c_{n}\|\bu\|_{\infty}^{2}H_{n} + 
\left|\I(\nabla^{n-1}\bom)\left[\nabla^{n-1}\left(\nabla\phi\times\Delta\nabla\phi\right)\right]\,dV\right|\non\\
&+& \left|\I (\nabla^{n-1}\bom)\left[\nabla^{n-1}\nabla^{\perp}\phi\right]\,dV\right|\,.
\eeq
Beginning with the third term on the right-hand side of Eq.~\eqref{hn1}, we obtain 
\bel{hn2}
\left|\I (\nabla^{n-1}\bom)\nabla^{n-1}\left(\nabla\phi\times\Delta\nabla\phi\right)\,dV\right|
\leq \|\nabla^{n-1}\bom\|_{2} \sum_{i=0}^{n-1}C^{n}_{i}\|\nabla^{i}(\nabla\phi)\|_{r}\|\nabla^{n+1-i}(\nabla\phi)\|_{s}\,.
\ee
Then, by using a Gagliardo-Nirenberg inequality,
\beq{hn3}
\|\nabla^{i}(\nabla\phi)\|_{r} &\leq& c\,\|\nabla^{n+1}(\nabla\phi)\|_{2}^{a} \|\nabla\phi\|_{\infty}^{1-a},\\
%\frac{1}{p} = \frac{i}{d} + a\left(\frac{1}{2} - \frac{n+1}{d}\right)\\
\|\nabla^{n+1-i}(\nabla\phi)\|_{s}
&\leq& c\,\|\nabla^{n+1}(\nabla\phi)\|_{2}^{b} \|\nabla\phi\|_{\infty}^{1-b}\,,
%\qquad\qquad \frac{1}{q} = \frac{n+1-i}{d} + b\left(\frac{1}{2} - \frac{n+1}{d}\right)\,.
\eeq
where $1/r + 1/s = 1/2$ and where
\beq{eq2}
\frac{1}{r} & = & \frac{i}{d} + a\left(\frac{1}{2} - \frac{n+1}{d}\right)\\
\frac{1}{s} & = & \frac{n+1-i}{d} + b\left(\frac{1}{2} - \frac{n+1}{d}\right)\,,
\eeq
we find that $a+b=1$. This yields
\beq{hn4}
\left|\I (\nabla^{n-1}\bom)\nabla^{n-1}\left(\nabla\phi\times\Delta\nabla\phi\right)dV\right|
%&=& \left|-\I \nabla^{n}\bom\,\nabla^{n-2}\left(\nabla\phi\times\Delta\nabla\phi\right)\,dV\right|\non\\
&\leq& c_{n}H_{n}^{1/2}P_{n+2}^{1/2}\|\nabla\phi\|_{\infty}\non\\
&\leq& P_{n+2} + \quart c_{n}H_{n}\|\nabla\phi\|_{\infty}^{2}\,.
\eeq
The last term on the right-hand side of Eq.~\eqref{hn1} is easily handled. Altogether we find 
\beq{hn5}
\shalf\dot{H}_{n} & \leq & -\shalf H_{n+1} + P_{n+2} + c_{n,3}\left(\|\bu\|_{\infty}^{2} + \|\nabla\phi\|_{\infty}^{2}\right)H_{n} 
+ \shalf H_{n} + \shalf P_{n}\,.
\eeq
\par\medskip\noindent
\textbf{Step 3\,:} Finally, by noting that $X_{n} =  P_{n+1} + H_{n}$, we use 
Eq.~\eqref{pn5} with $n \to n+1$ to obtain
\beq{hn6}
\shalf \dot{X}_{n} &\leq& - \shalf P_{n+3} + \threehalves P_{n+2} + 
c_{n,1}\left(\shalf\|\phi\|_{\infty}^{4} + \|\bu\|_{\infty}^{2}\right)P_{n+1} 
+ c_{n,2}H_{n}\|\nabla\phi\|_{\infty}^{2}\non\\
&-& \shalf H_{n+1} + P_{n+2} + c_{n,3}\left(\|\bu\|_{\infty}^{2} + \|\nabla\phi\|_{\infty}^{2}\right)H_{n} 
+ \shalf H_{n} + \shalf P_{n}\non\\
&\leq& - \shalf P_{n+3} - \shalf H_{n+1} + \fivehalves P_{n+2} +
c_{n,4}\left(\shalf\|\phi\|_{\infty}^{4} + \|\bu\|_{\infty}^{2} + 
\|\nabla\phi\|_{\infty}^{2}\right)X_{n} + \shalf H_{n} + \shalf P_{n}\,.
\eeq
By using $P_{n+2} \leq P_{n+3}^{1/2}P_{n+1}^{1/2} \leq (\varepsilon/2) P_{n+3} + 
(1/2\varepsilon)P_{n+1}$, with $\varepsilon$ chosen as $\varepsilon = \onefifth$, we have (with $P_{n} \leq P_{n+1}$)
\bel{hn7}
\shalf \dot{X}_{n} \leq - \quart P_{n+3} - \shalf H_{n+1} + 
c_{n,4}\left(\|\nabla\phi\|_{\infty}^{2} + \shalf\|\phi\|_{\infty}^{4} + \|\bu\|_{\infty}^{2} + \shalf\right)X_{n}\,.
\ee
We note that $\phi$ is a mean-zero function on a unit periodic domain, so $\|\phi\|_{\infty}\leq \|\nabla\phi\|_{\infty}$. 
Then we can write
\beq{hn8}
c_{n,4}\left(\|\nabla\phi\|_{\infty}^{2} + \shalf\|\phi\|_{\infty}^{4} + \|\bu\|_{\infty}^{2} + \shalf\right) 
&=&  c_{n,4}\left(\|\nabla\phi\|_{\infty}^{2} + \shalf\left(\|\phi\|_{\infty}^{2}-1\right)^{2}+\|\bu\|_{\infty}^{2} 
+ \|\phi\|_{\infty}^{2}\right)\non\\
&\leq& 2c_{n,4}\left(\|\nabla\phi\|_{\infty}^{2} + \shalf\left(\|\phi\|_{\infty}^{2}-1\right)^{2} + \|\bu\|_{\infty}^{2}\right)\,.
\eeq
By dropping the negative terms, Eq.~\eqref{hn7} turns into 
\bel{hn8}
\quart \dot{X}_{n} \leq c_{n,4}E_{\infty} X_{n}\,,
\ee
where $E_{\infty}$ is defined in Eq.~\eqref{Einf}. By integrating over $[0,\,T^{*}]$, we obtain
\bel{Xnexp}
%X_{n}(T^{*}) &\leq& x %c_{n,5}X_{n}(0)\exp \int_{0}^{T^{*}}E_{\infty}(\tau)\,d\tau\,.
X_{n}(T^{*}) \leq c_{n,5}X_{n}(0)\exp \int_{0}^{T^{*}}E_{\infty}(\tau)\,d\tau\,.
\ee
The assumption that the time integral is finite implies that $X_{n}(T^{*}) < \infty$, which contradicts the statement 
in the Theorem that solutions first lose regularity at $T^{*}$. \hfil$\blacksquare$
\end{widetext}


\begin{thebibliography}{}

\bibitem{navier1823memoire} C.L.M.H. Navier, ``Memoire sur les lois du mouvement des fluides'', 
Mem. Acad. Sci. Inst. France, \textbf{6}, 389--440 (1822)\,; G.G. Stokes, \textit{Mathematical and 
Physical Papers}, Vol. \textbf{1} (Cambridge University Press, Cambridge, UK, 1880).

\bibitem{leray1934mouvement} J. Leray, %Sur le mouvement d'un liquide visqueux emplissant l'espace, 
Acta Math., \textbf{63}, 193–248 (1934). 

\bibitem{batchelor1967introduction} G.K. Batchelor, \textit{An Introduction to Fluid Mechanics} 
(Cambridge University Press, Cambridge, UK, 2000)\,; http://dx.doi.org/10.1017/CBO9780511800955.

\bibitem{fefferman2006existence} %Geometric constraints on potentially singular solutions for the 3-D Euler equations
P. Constantin, C. Fefferman and A.J. Majda, Comm. PDEs \textbf{21} (3-4) (1996). 
\bibitem{foias2001navier} C. Foias, O. Manley, R. Rosa and R. Temam, \textit{Navier-Stokes Equations and Turbulence}, 
(Cambridge University Press, Cambridge, UK, 2001) 

\bibitem{doering1995applied} C.R. Doering and J.D. Gibbon, \textit{Applied Analysis of the Navier-Stokes Equations} 
(Cambridge University Press, Cambridge, UK, 2004). 

\bibitem{cahn1958free} J.W. Cahn and J.E. Hilliard, %Free energy of a non-uniform system. I. Interfacial free energy, 
J. Chem. Phys \textbf{28}, 258--267 (1958).

\bibitem{chaikin2000principles}
P.M. Chaikin and T.C. Lubensky, \textit{Principles of Condensed Matter Physics}  
(Cambridge University Press, Cambridge, UK, reprint edition 2000) .

\bibitem{hohenberg1977theory} P.C. Hohenberg and B.I. Halperin, Rev. Mod. Phys. \textbf{49} 435--480 (1977).

\bibitem{lifshitz1961kinetics} I.M. Lifshitz and V. V. Slyozov, J. Phys. Chem. Solids \textbf{19}, 35--50 (1959)\,; 
H. Furukawa, Phys. Rev. A \textbf{31}, 1103--1108 (1985)\,; E. D. Siggia, Phys. Rev. A \textbf{20}, 595--605 (1979).

\bibitem{gunton1983m} J.D. Gunton, M. San Miguel, and P.S. Sahni, in \textit{Phase Transitions and 
Critical Phenomena}, eds. C. Domb and J.L. Lebowitz, \textbf{8} (Academic, London, 1983).

\bibitem{bray1994theory} A. J. Bray, Adv. Phys., \textbf{43}, 357--459, 1994.


\bibitem{muskat1} A. Castro, D. Cordoba, Ch. Fefferman, F.Gancedo and M. L\'{o}pez-Fern\'{a}ndez, 
% "Rayleigh-Taylor breakdown for the Muskat problem with applications to water waves".
Ann. Math. \textbf{175}(2), 909--948 (2012). 

\bibitem{muskat2} A. Castro, D. Cordoba, Ch. Fefferman and F. Gancedo, 
% "Breakdown of smoothness for the Muskat problem"
Arch. Ration. Mech. Anal., \textbf{208}, 805--909 (2013).


\bibitem{puri2009kinetics} S. Puri, in \textit{Kinetics of Phase Transitions}, eds. S. Puri and V. Wadhawan, 
\textbf{6}, p. 437, (CRC Press, Boca Raton, US, 2009).

\bibitem{lothe1962reconsiderations} J. Lothe and G.M. Pound, J. Chem. Phys. \textbf{36}, 2080--2084 (1962).

\bibitem{onuki2002phase} A. Onuki, \textit{Phase Transition Dynamics} (Cambridge University Press, Cambridge, UK, 2002).

\bibitem{badalassi2003computation}
V.E. Badalassi, H.D. Ceniceros, and S. Banerjee, J. Comput. Phys. \textbf{190}, 371--397 (2003).

\bibitem{perlekar2014spinodal}
P. Perlekar, R. Benzi, H.J.H. Clercx, D.R. Nelson and F. Toschi, Phys. Rev. Lett, \textbf{112}, 014502 (2014).

\bibitem{cahn1961spinodal} J.W. Cahn, Acta Metallurgica \textbf{9}, %On spinodal decomposition, 
795--801 (1961).

\bibitem{berti2005turbulence}
S. Berti, G. Boffetta, M. Cencini and A. Vulpiani, Phys. Rev. Lett. \textbf{95}, 224501 (2005)\,; 
A.J. Wagner and J.M. Yeomans, Phys. Rev. Lett. \textbf{80}, 1429 (1998)\,; 
V.M. Kendon, Phys. Rev. E, \textbf{61}, R6071 (R) (2000)\,;
V.M. Kendon, M.E. Cates, I. Pagonabarraga, J.C. Desplat, P. Blandon, J. Fluid Mech., \textbf{440}, 147--203 (2001).

\bibitem{boffetta2012two} G. Boffetta and R. Ecke, Annu. Rev Fluid Mech. \textbf{44}, 427--451 (2012)\,; 
R. Pandit, P. Perlekar, and S. S. Ray, Pramana-Journal of Physics, \textbf{73}, 157--191 (2009).

\bibitem{cabot2006reynolds} W.H. Cabot, and A.W. Cook, Nat. Phys. \textbf{2} (8), 562--568 (2006).

\bibitem{celani2009phase} A. Celani, A. Mazzino, P. Muratore--Ginanneschi and L.
Vozella, J.Fluid Mech., \textbf{622}, 115--134 (2009).

\bibitem{nairita2016}
N. Pal, P. Perlekar, A. Gupta, and R. Pandit, Phys. Rev. E \textbf{93}, 063115 (2016).

\bibitem{shardt2013simulations} %{\color{red}Coalesce??}
O. Shardt, S. Mitra, J. Derksen, Langmuir {\bf 30} 14416--14426 (2014). 

\bibitem{LBM} A. Gupta and M. Sbragaglia, Phys. Rev. E \textbf{90} 023305 (2014).
\bibitem{scarbolo2013turbulence} L. Scarbolo and A. Soldati, J. Turb. \textbf{14}, 11 (2013). 

\bibitem{yue2004diffuse} P. Yue, J.J. Feng, C. Liu, and J. Shen, J. Fluid Mech. \textbf{515}, 293--317 (2004).

\bibitem{scarbolo2011phase} L. Scarbolo, D. Molin and A. Soldati, APS Div. Fluid Dynamics Abstracts.  \textbf{1}, 4002 (2011).

\bibitem{abels2009longtime} H. Abels, ``Longtime behavior of solutions of a Navier–Stokes/Cahn–Hilliard system'', in\,: 
Proceedings of the Conference \textit{Nonlocal and Abstract Parabolic Equations and Their Applications}, 
Bedlewo, in\,: Banach Center Publ., Polish Acad. Sci., \textbf{86}, 9--19 (2009).

\bibitem{gal2010asymptotic} C.G. Gal and M. Graselli, Ann. I. H. Poincar\'e -- AN, \textbf{27}, 401--436 (2010).

\bibitem{constantin1996geometric}
P. Constantin and C. Fefferman, Indiana Univ. Math. J. {\bf 42} (1993).

\bibitem{elliott1986cahn} C.M. Elliott and Z. Songmu, Arch. Rat. Mech. Anal. \textbf{96} (4), 339--357 (1986).

\bibitem{majda2001vorticity} A.J. Majda and A. Bertozzi, \textit{Vorticity and incompressible flow}, Cambridge Texts in 
Applied Mathematics (No. 27), (Cambridge University Press, Cambridge, UK, 2001).

\bibitem{gibbon2008three} J.D. Gibbon, ``The three dimensional Euler equations\,: 
how much do we know?'' Proc. of ``Euler Equations 250 years on'', Aussois 
June 2007, Physica D, \textbf{237}, 1894--1904 (2008).

\bibitem{bardos2007euler} C. Bardos and E.S. Titi ``Euler equations of incompressible ideal fluids'', Russ. Math. Surv. 
\textbf{62:3} 409--451 (2007).


\bibitem{Tran2016} C. V. Tran and X. Yu, 
% Pressure moderation and effective pressure in Navier–Stokes flows
Nonlinearity, \textbf{29}, 2990--3005 (2016). 

\bibitem{beale1984remarks} J.T. Beale, T. Kato, and A.J. Majda, ``Remarks on the breakdown of smooth solutions 
for the 3D Euler equations'' Commun. Math. Phys., \textbf{94} 61--66 (1984).

\bibitem{donzis2013vorticity} D.A. Donzis, J.D. Gibbon, A. Gupta, R.M. Kerr, R. Pandit, and
D. Vincenzi,  J. Fluid Mech. \textbf{732}, 316--331 (2013).

\bibitem{gibbon2014regimes} J.D. Gibbon, D.A. Donzis, A. Gupta, R.M. Kerr, R. Pandit and D. Vincenzi, 
``Regimes of nonlinear depletion and regularity in the 3D Navier-Stokes equations,'' Nonlinearity 
\textbf{27}, 1--19 (2014).

\bibitem{gibbon2016high} J.D. Gibbon, IMA J. Appl. Math., \textbf{81}(2) 308-320 (2015). 

\bibitem{gibbon2016depletion} J.D. Gibbon, A. Gupta, G. Krstulovic, R. Pandit, H. Politano, Y. Ponty, A. Pouquet, G. Sahoo, and
J. Stawartz, Phys. Rev. E, \textbf{93}, 043104 (2016).

\bibitem{petrasso1994rayleigh} R.D. Petrasso, Nat. Phys. {\bf 367} (6460), 217--218 (1994).

\bibitem{taleyarkhan2002evidence} R.P. Taleyarkhan, Science \textbf{295}, 1868--1873 (2002).

\bibitem{munk1998abyssal} W. Munk and C. Wunsch, Deep-Sea Res. I \textbf{45} (12), 1977--2010 (1998).

\bibitem{bhatnagar2014universal} A. Bhatnagar, A. Gupta, D. Mitra, P. Perlekar and R. Pandit, arXiv preprint arXiv:1412.2686 (2014).

\bibitem{waddell2001experimental} J.T. Waddell, C.E. Niederhaus, and J.W. Jacobs, Phys. Fluids \textbf{13}, 1263--1273 (2001).

\bibitem{ramaprabhu2006limits} P. Ramaprabhu, G. Dimonte, Y.N. Young, A.C. Calder and B.  Fryxell, Phys. Rev. E \textbf{74}, 066308 (2006).

\bibitem{dalziel2008mixing} S.B. Dalziel \textit{et al.}, Phys. Fluids \textbf{20}, 065106 (2008).

\bibitem{rao2016nonlinear} P. Rao, C.P. Caulfield, and J.D. Gibbon, ``Nonlinear effects in buoyancy-driven variable
density turbulence'' http://arxiv.org/pdf/1601.03445.pdf (2016).

\bibitem{livescu2013numerical} D. Livescu, ``Numerical simulations of two-fluid mixing at large density ratios and applications 
to the Rayleigh-Taylor instability'', Phil. Trans. R. Soc. A. \textbf{371}, 20120185 (2013).
\bibitem{suppmat}
See the Supplemental Material at \burl{https://www.youtube.com/watch?v=VZXMJceTKWw&feature=youtu.be} for the video sequence showing
isosurface plots of the $\phi$-field for parameters of our simulations {\tt T1}. 
Here kinematic viscosity $=1.167 \times 10^{-3}$. For this video, we use $10$
frames per second with each frame separated from the succeeding frame by $200 \delta t$, where
$\delta t=10^{-3}$.

%\bibitem{fefferman2006existence} %Geometric constraints on potentially singular solutions for the 3-D Euler equations
%P. Constantin, C. Fefferman and A.J. Majda, Comm. PDEs \textbf{21} (3-4) (1996). 

\end{thebibliography}
\end{document}